\documentclass[9pt,twocolumn]{extarticle}

\usepackage[sort&compress,numbers]{natbib}
\usepackage{fancyvrb}
\usepackage{xspace}
\usepackage{comment}
\usepackage{libertine}
\usepackage{graphicx}
\usepackage[font={small}]{caption}
\usepackage{mdwlist}
\usepackage{booktabs}
\usepackage[T1]{fontenc}
\usepackage[usenames, dvipsnames]{xcolor}
\usepackage{censor}
\usepackage{tabularx}
\usepackage{fullpage}
\usepackage{authblk}

\usepackage{fancyhdr}
\date{}

\PassOptionsToPackage{hyphens}{url}\usepackage{hyperref}

\newcommand{\eat}[1]{}
\newcommand{\pages}[1]{}
\newcommand{\confirmed}[1]{#1} 

\newcommand{\cmark}{\ding{51}}%
\newcommand{\xmark}{\ding{55}}%
\newcommand{\Paragraph}[1]{\vspace{.05in}\noindent{\bf #1.}\quad}
\newcommand{\mydate}[1]{\mbox{#1}}
\newcommand{\html}[1]{\mbox{\tt #1}}
\newcommand{\Section}[1]{\vspace{0mm}\section{#1}}
\newcommand{\text}[1]{\ensuremath\textrm{#1}}

\newcommand{\mytitle}{An Extensive Evaluation of the Internet's Open Proxies}

\hypersetup{%
  pdftitle={\mytitle},
  pdfauthor={Akshaya Mani, Tavish Vaidya, David Dworken, and Micah Sherr}
  pdfdisplaydoctitle=true, 
  bookmarksnumbered,
  pdfstartview={FitH},
  colorlinks,
  citecolor=black,
  filecolor=black,
  linkcolor=black,
  urlcolor=black,
  breaklinks=true,
  hypertexnames=false
}

\begin{document}

\title{\mytitle}
\author[1]{Akshaya Mani\thanks{Co-first authors.}}
\author[1]{Tavish Vaidya\thanks{Co-first authors.}} 
\author[2]{David Dworken}
\author[1]{Micah Sherr}
\affil[1]{Georgetown University}
\affil[2]{Northeastern University}

\maketitle
\thispagestyle{fancy}
\begin{abstract}

  Open proxies forward traffic on behalf of any Internet user.  Listed
  on open proxy aggregator sites, they are often used to bypass
  geographic region restrictions or circumvent censorship.  Open
  proxies sometimes also provide a weak form of anonymity by
  concealing the requestor's IP address.  

  To better understand their behavior and performance, we conducted a
  comprehensive study of open proxies, encompassing more than 107,000
  listed open proxies and 13M proxy requests over a 50 day period.
  While previous studies have focused on malicious open proxies'
  manipulation of HTML content to insert/modify ads, we provide a more broad 
  study that examines the availability, success rates, diversity, and also 
  (mis)behavior of proxies.

  Our results show that listed open proxies suffer poor
  availability---more than 92\% of open proxies that appear on
  aggregator sites are unresponsive to proxy requests.  Much more
  troubling, we find numerous examples of malicious open proxies
  in which HTML content is manipulated to mine cryptocurrency (that
  is, cryptojacking).  We additionally detect TLS man-in-the-middle
  (MitM) attacks, and discover numerous instances in which binaries
  fetched through proxies were modified to include remote access
  trojans and other forms of malware.
  As a point of comparison, we conduct and discuss a similar
  measurement study of the behavior of Tor exit relays.  We find no
  instances in which Tor relays performed TLS MitM or manipulated
  content, suggesting that Tor offers a far more reliable and safe
  form of proxied communication.
\end{abstract}

\Section{Introduction \pages{1.25}}


{\em Open proxy servers} are unrestricted proxies that allow access
from any Internet user.  Open proxies are fairly prevalent on the
Internet, and several websites are devoted to  maintaining large lists of
available open proxies.
In contrast to VPNs~\cite{ferguson2000cryptographic} and some (but not
all) anonymity systems, open proxies are generally easy for users to
use, requiring only minimal configuration changes (e.g., adjusting
``network settings'') rather than the installation of new software.

As surveyed in this paper, there is a variety of types of open
proxies, with differing capabilities.  This leads to
variation in how proxies are used in practice.  Clearly, open
proxies have been (mis)used for malicious purposes, including (but not
limited to) sending spam, injecting ads, and serving as stepping stones for various
attacks~\cite{dark-side-of-web,tsirantonakis2018large}.
But open proxies have also been used for
far less nefarious reasons, including circumventing censorship
efforts.  More generally, they permit accessing otherwise inaccessible
information and, controversially, can be used to bypass regional
content filters (e.g., to watch a sporting event that, due to
licensing restrictions, would otherwise not be viewable).  Importantly, along with VPNs,
proxies have been suggested as a means of enhancing Internet privacy
and protecting browsing history in the face of the recent expansion
of data collection rights by U.S.~Internet service providers~\cite{privacy-law}.

An interesting related question, and one that unfortunately has not
undergone rigorous study, is {\em why} the owners of open proxy
servers run such services.  In some instances, the proxies are due to
misconfiguration or even compromise. 
There is a rich history of accidental open proxies, particularly for
open proxies that operate as SMTP relays~\cite{spam-relays}. Still, many
proxies are operated by choice.  Although the rationales here are not
well-understood, we posit that some operators may run proxies as a
political statement about privacy and the freedom to access
information.

A more menacing motivation for running a proxy is that it provides its
operator with an expanded view of network traffic.  Proxy operators
are well situated to eavesdrop on communication, perform
man-in-the-middle (MitM) attacks, and monitize their service by
injecting spurious ads~\cite{tsirantonakis2018large}. We note that
while it is relatively difficult to operate a backbone Internet router
and monitor traffic, it is not difficult to act as an open proxy and
increase one's view of others' communication.

We are of course not the first to suggest that open proxies should be
treated with a healthy supply of skepticism, due in large part to the
ease in which proxies can be instantiated and configured to eavesdrop
and manipulate communication.  Most noteworthy is the recent study of
open proxies by Tsirantonakis et al.~\cite{tsirantonakis2018large}.
There, the authors explore the extent to which open proxies modify
proxied HTML pages to inject ads, collect user information, and
redirect users to malware-containing pages.  Similar to Tsirantonakis
et al.'s study, we also consider the modification of retrieved
webpages.  This paper compliments their work by conducting a large
and more broad study of the Internet's open proxies.


We perform an extensive study of the open proxy ecosystem,
consisting of more than 107,000 publicly listed open
 proxies.  We find that the vast
majority (92\%) of open proxies that are publicly listed on aggregator
websites are either unavailable or otherwise do {\em not} allow proxy
traffic.  Surprisingly, the open proxies that do allow proxy traffic
have very little geographic or network diversity: five countries
account for nearly 60\% of the Internet's working open proxies, and
41\% of such proxies are hosted by just ten autonomous systems (ASes).
We also evaluate the performance of the proxies, and find a wide
distribution of effective goodputs.

To understand the security implications of relaying traffic through
open proxies, we present a series of experiments designed to
understand a broad range of proxy behaviors.  We compare
the results of traffic received via direct communication with traffic
that traverses through proxies.
  While our results indicate that the
majority of open proxies seemingly operate correctly (that is, they
forward the traffic unimpeded), we find a surprisingly large number
of instances of misbehavior.  In particular, as we discuss in more
detail in what follows, we discover many instances in which proxies
manipulate HTML content, not only to insert ads, but also to mine cryptocurrency
(i.e., {\em cryptojacking}).  Our study also discloses numerous attacks in
which proxies return trojan Windows executables and remote access
trojans (RATs), and conduct TLS MitM attacks.

As a point of comparison, we compare the level of manipulation
observed using open proxies to that when content is fetched over
Tor~\cite{tor}.  Over the course of our
nearly \confirmed{monthlong} experiment in which we fetched files from every
available Tor exit relay, \confirmed{we found no instances in which
  the requested content was manipulated}.  Our results suggest that Tor offers a more
reliable and trustworthy form of proxied communication.



In summary, this paper describes a large-scale  and broad study of the
Internet's open proxies that (i)~examines the open proxy ecosystem,
(ii)~measures open proxy performance, (iii)~measures the prevelance of
previously undisclosed
 attacks, and (iv)~relates its findings to a similar study (which we
 conduct) of Tor exit relay behavior.  Our results
demonstrate that misbehavior abounds on the Internet's open proxies
and that the use of such proxies carries substantial risk.

\Section{Background \pages{0.75}}
\label{sec:background}


The Socket Secure (SOCKS) protocol~\cite{rfc1928-socksv5} was
introduced in 1992 as a means to ease the configuration of network
firewalls~\cite{socks}.  In SOCKS, a server listens for incoming TCP
connections from a client.  Once connected, a client can then tunnel
TCP and/or UDP traffic through the SOCKS proxy to its chosen
destination(s).  Since SOCKS forwards arbitrary TCP and UDP traffic,
clients can use end-to-end (e2e) security protocols (in particular, TLS/SSL)
through SOCKS proxies.  Currently, SOCKSv4 and SOCKSv5 servers are
both deployed on the Internet, with the latter adding authentication
features that enforce access control.  In this paper, we
do not distinguish between SOCKSv4 and SOCKSv5, and use the general
term {\bf SOCKS proxies} to describe proxies running either version.  

More commonly (see \S\ref{sec:performance}), open proxies use HTTP as
a transport mechanism.  This has the benefit of supporting clients
whose local firewall policies prohibit non-HTTP traffic (e.g., as is
sometimes the case on corporate networks).\footnote{Although, as we
  report in \S\ref{sec:performance}, most proxies that use HTTP as a
  transport mechanism do not listen on ports 80 or 443, potentially nullifying this
  advantage.}

We distinguish between two types of proxies that rely
on HTTP.  The first, which we refer to simply as {\bf HTTP proxies},
allows clients to specify a fully-qualified URL---with any domain
name---as part of a HTTP GET request to the proxy.
The proxy parses the GET URL and if the requested URL
is on a different host, the proxy makes its own request to fetch the
URL and forwards the response back to the client.  Conceptually, a
HTTP proxy is a web server that serves pages that are hosted
elsewhere.

A significant disadvantage of HTTP proxies is that it is incompatible
with e2e security protocols such as TLS, since it is the HTTP proxy
(not the user's browser) that initiates the connection to the proxied
URL.  The inability to support fetching HTTPS URLs increasingly limits
the use of HTTP proxies due to the growing adoption rate of
HTTPS~\cite{felt2017measuring} on the web.

In contrast, {\bf CONNECT proxies} use the HTTP CONNECT \protect\linebreak
method~\cite{rfc2817-connect} to establish e2e tunnels between the
client (i.e., the browser) and the destination webserver.  CONNECT proxies allow the client
to specify a host and port.  The proxy then initiates a TCP connection
to the specified target 
and then transparently forwards all future TCP data from the client to the
destination, and vice versa.  Importantly, CONNECT proxies forward
traffic at the TCP level (layer~4) and are not limited to any particular
application-layer protocol.  CONNECT proxies
support TLS/HTTPS connections since the browser can effectively
communicate unencumbered to the destination webserver and perform an
ordinary TLS handshake.  Example protocol interactions for HTTP and
CONNECT proxies are provided in Figures~\ref{fig:http}
and~\ref{fig:connect} in
Appendix~\ref{app:examples}.

The network addresses (IPs and ports) of open proxies are listed
on {\bf proxy aggregator sites}.  These sites also sometimes
use the term {\em transparent}, meaning that the proxy supports at
least e2e TCP tunneling; SOCKS and CONNECT proxies meet this
criterion.  As another dimension, aggregator sites also sometimes
categorize proxies as {\em anonymous proxies}.  These proxies purportedly do not
reveal the client's IP address to the destination.  
We explore the degree to which
open proxies provide anonymity in~\S\ref{sec:performance}.

\Section{Related Work \pages{0.5}}
\label{sec:related}

We are not the first to investigate the use of proxies on the Internet.  Weaver et
al.~\cite{Weaver2014} performed a measurement study to detect the presence of {\em transparent} HTTP
proxies on connections from clients to servers.  They found that 14\% of tested connections were
proxied somewhere along the network path between the client and destination.  Our work focuses on
the {\em intentional} use of freely available proxies and gathers evidence of misbehavior
by such proxies.

Scott et al.~\cite{scott2015understanding} provide insights about the use, distribution, and traffic
characteristics of open proxies on the Internet.  However, they do not focus on detecting any
malicious behavior by open proxies, which is a core focus of this work.

Other work has investigated the misuse of proxy servers.  Of note, Jessen et al.~\cite{steding2008using}
deploy low-interaction honeypots to explore the (ab)use of open proxies
to send spam. Researchers have also examined the abuse of the Codeen CDN network~\cite{codeen} by looking at the
requests that were forwarded by the network and mitigation strategies employed to minimize the
forwarding of malicious requests~\cite{dark-side-of-web, wang2004reliability}. While we also
investigate misbehavior, we focus on uncovering malicious behavior by open proxies
rather than  malicious use of proxies by users.

Tyson et al.~\cite{Tyson:2017:EHH:3038912.3052571} study HTTP header
manipulation on the Internet by various open proxies and discuss various observed factors affecting
HTTP header manipulation. Huang et al.~\cite{huang2017middleboxes}
show that ASes sometimes use middleboxes 
that interfere with HTTP traffic and inject HTTP headers across a wide range of networks including mobile and data centers.
Durumeric et al.~\cite{durumeric2017security} examine the degree of HTTPS
interception by middleboxes and argue that such interception significantly reduces security.  Our
work has  greater scope, and  focuses not only on  header manipulation or HTTPS
interception, but also uncovers several other forms of suspicious behavior exhibited by open
proxies. Waked et al.~\cite{waked2018intercept} show that commercially used TLS middleboxes do not perform
sufficient validation checks on SSL certificates and expose their clients to trivial attacks. \confirmed{We
also found that some open proxies exhibit similar behavior.}

Previous work has also looked at ad injection in HTML content~\cite{reis2008detecting,
thomas2015ad,chung2016tunneling}. Similarly, we look at content manipulation by
open proxies, but consider other forms of misbehavior beyond ad injection.

Most related to this work is the recent study by Tsirantonakis et al.~\cite{tsirantonakis2018large}
that also looks at content manipulation by examining roughly 66K open HTTP proxies on the Internet. Their work provides an
in-depth analysis of different types of malicious behavior exhibited by proxies by injecting Javascript code.
Their analysis shows that proxies inject Javascript aimed at tracking users, stealing
user information, fingerprinting browsers, and replacing  ads.
Complementary to their work, we provide an in-breadth analysis of 107K
unique open proxies by evaluating their availability, performance, and
 behavior across a large spectrum of potential attacks.
We look at
manipulation of content by open proxies, not only through the
injection of Javascript, but also by modification of different types of
requested content (binary files, etc.) and through TLS
MitM. Additionally, unlike existing work, we
evaluate the behavior of open proxies based on different client locations.
Laudably, Tsirantonakis et al.~\cite{tsirantonakis2018large} built a service to  detect malicious
open proxies on a daily basis and to publicly report the proxies
deemed unsafe. We compare our findings using this service. 
As an alternative to open proxies, we compare the performance of open proxies
to Tor exit relays (\S\ref{sec:tor}), and argue that Tor provides a
safer means of proxied communication.




\Section{Methodology \& Experimental Setup \pages{1}}
\label{sec:overview}

Our study of the Internet's open proxy servers has two main goals:
(i)~to measure the proxies' availability, composition, and performance and
(ii)~to assess the degree to which proxies exhibit malicious behavior.
We begin by describing the methodology used to perform our study.

We conducted our study over a 50 day period, beginning on
\mydate{2018-04-12} and ending on \mydate{2018-05-31}.  Our
measurement apparatus, described next, was installed on 16 locations
(listed in Table~\ref{tbl:clients} in Appendix~\ref{app:clients});
these included 15 geographically diverse AWS regions and an
installation at our local institution.  We term each instance a {\em
  client location}.  Having multiple client locations allows us to
determine whether proxies behave differently based on the network
and/or geographic locations of the requesting client.

\begin{table}[t]
\centering
\begin{small}
\begin{tabular}{lr}
    \toprule
    {\bf Proxy Aggregator} & {\bf Proxies} \\
    \midrule
  clarketm~\cite{clarketm}  & 6,343 \\
  multiproxy.org (all)~\cite{multiproxy1}  & 1,524 \\
  multiproxy.org (anon)~\cite{multiproxy2}  & 373 \\
  NordVPN~\cite{nordvpn} & 29,194 \\
  ProxyBroker~\cite{proxybroker} & 73,905 \\
  workingproxies.org~\cite{workingproxies} & 1,250 \\
  xroxy~\cite{xroxy} & 345 \\
  \midrule
  Total (unique proxies) & 107,034 \\ 
\bottomrule
\end{tabular}
\end{small}
\caption{Sources of open proxies. A given proxy may be listed by more than one aggregator.}
\label{tbl:sources}
\end{table}

From each client location, an automated process performed the
following steps once per day:
\begin{enumerate}
\item {\bf Populate:}\quad We collect and combine lists of advertised
  proxies from a number of proxy aggregator sites.  We augment this
  list with the results of running
  \mbox{ProxyBroker~\cite{proxybroker}}, an open source tool for
  finding open proxies.  In all cases, proxies were listed by
  aggregator sites as tuples containing an IPv4 address and a TCP port
  number. The complete list of sources of proxies is listed in
  Table~\ref{tbl:sources}.  We emphasize
  that the list of proxies is (re)fetched daily since, as we show in
  \S\ref{sec:performance}, open proxies are subject to high levels of
  churn.

\item {\bf Classify:}\quad Next, from each client location, we attempt
  to connect to each proxy and, if successful, determine whether the
  proxy is a HTTP, CONNECT, or SOCKS proxy.

\item {\bf Fetch:}\quad Finally, we request several files (URLs) from
  the set of proxies that we were successfully able to classify.

\end{enumerate}

In more detail, during the Fetch step, we retrieve the following files
from each proxy that we were able to classify, using unencrypted HTTP
connections: an HTML page, a Flash object (.swf), a Windows executable
(.exe), a JPEG image, a ZIP file, a Windows batch (.bat) file, a
Linux/UNIX shell script (.sh), and a Java JAR archive.  With the
exception of the .exe file (explained in more detail in
\S\ref{sec:files}), the URLs are hosted on web servers at our
institution.

For CONNECT and SOCKS proxies (i.e., the proxies that support TLS), we
also request files over HTTPS from a properly configured (i.e.,
with a valid certificate) web server running at our institution.  We
additionally request HTML files from \url{https://revoked.badssl.com/}
and \url{https://self-signed.badssl.com/} which, respectively, use
revoked and self-signed certificates.  The rationale for fetching
content from sites with invalid certificates is discussed in
\S\ref{sec:ssl}.
In all cases, we set the User-Agent HTTP request header to match that
of Google Chrome version 62.0.3202.94 running on \mbox{Mac OS~X}.

For each proxy request, we record whether the request completed.  If
we received a response from the proxy, we also record the HTTP status
code and response string (e.g., ``{\tt 200~OK}'') returned by the proxy, the
size of the response, the content of the response, the MIME-type of
the response (as determined by filemagic), the time-to-last-byte
(TTLB) for receiving the response, the HTTP response headers, and (in
the case of HTTPS requests) the certificate received.

Throughout the remainder of this paper, we use the term {\bf expected
  content} to refer to the {\em correct} contents of the file and
{\bf unexpected content} to refer to content returned by a proxy
that does not match the file indicated by the requested URL.  A
correctly functioning open proxy should thus return the expected
content.  As we
explore in more detail below, unexpected content does not necessarily
indicate malicious behavior.  For instance, unexpected content could
be an HTML page indicating that the proxy is misconfigured or that the
proxy requires user authentication.

We remark that a weakness of our study, and one that we share with
studies that examine similar (mis)behavior in anonymity
and proxy services~\cite{tsirantonakis2018large,SpoiledOnions,chakravarty2011detecting}, is that we cannot
easily differentiate between manipulation that occurs at a proxy and
manipulation that occurs somewhere along the path between the proxy
and the destination.  
Our findings of
misbehavior can be viewed as indications that a proxy should not be
used, either because it was itself malicious or because its network
location makes traffic routed through it  routinely vulnerable.

\Section{Proxy Availability \& Performance \pages{1}}
\label{sec:performance}

\begin{figure*}[t]
  \centering
  \begin{minipage}{.32\linewidth}
    \centering
    \includegraphics[width=1.0\linewidth]{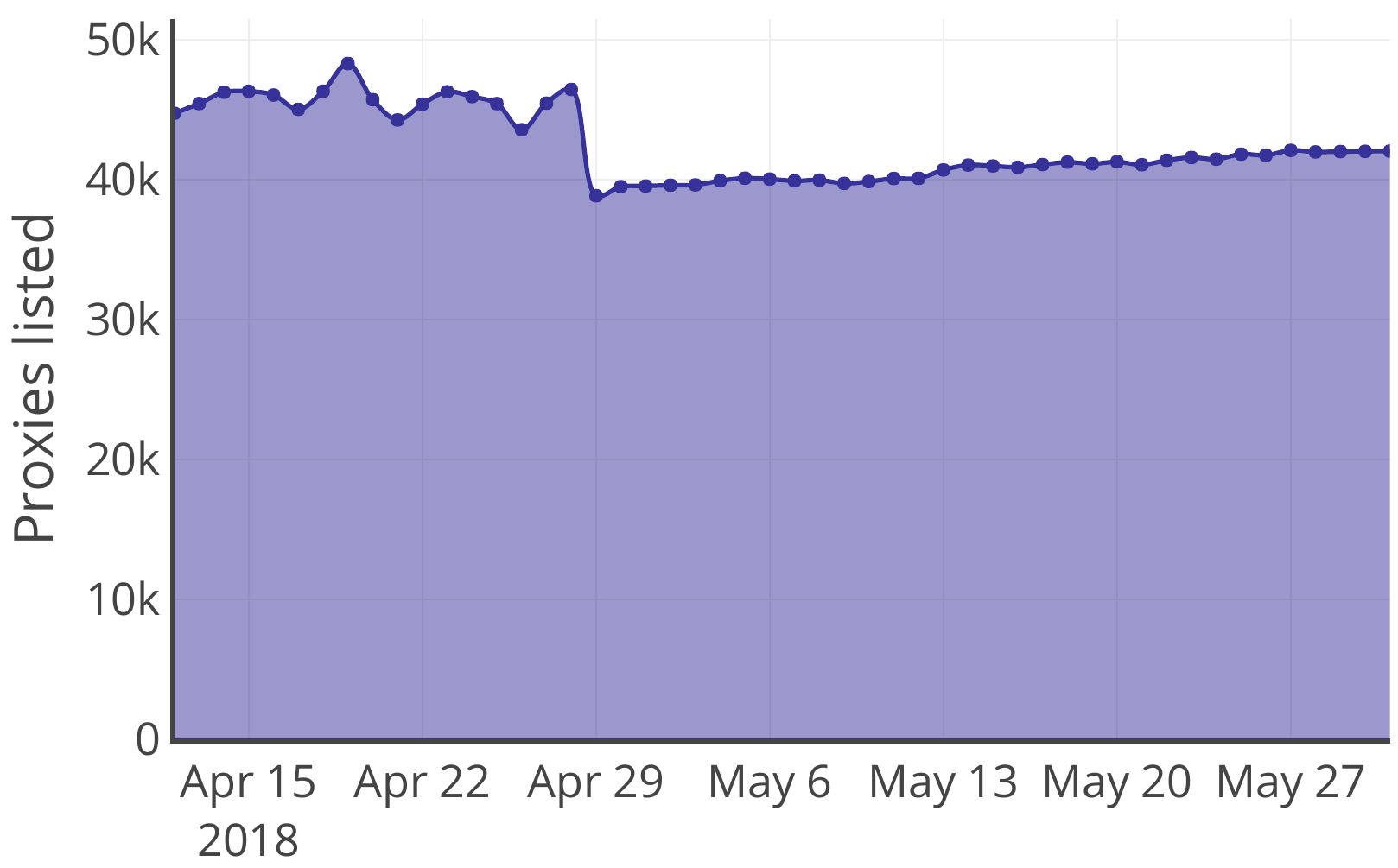}
    \caption{Number of unique open proxies listed on aggregator sites, over time.}
    \label{fig:numlisted}
  \end{minipage}
  \hfill
  \begin{minipage}{.32\linewidth}
    \centering
    \includegraphics[width=1.0\linewidth]{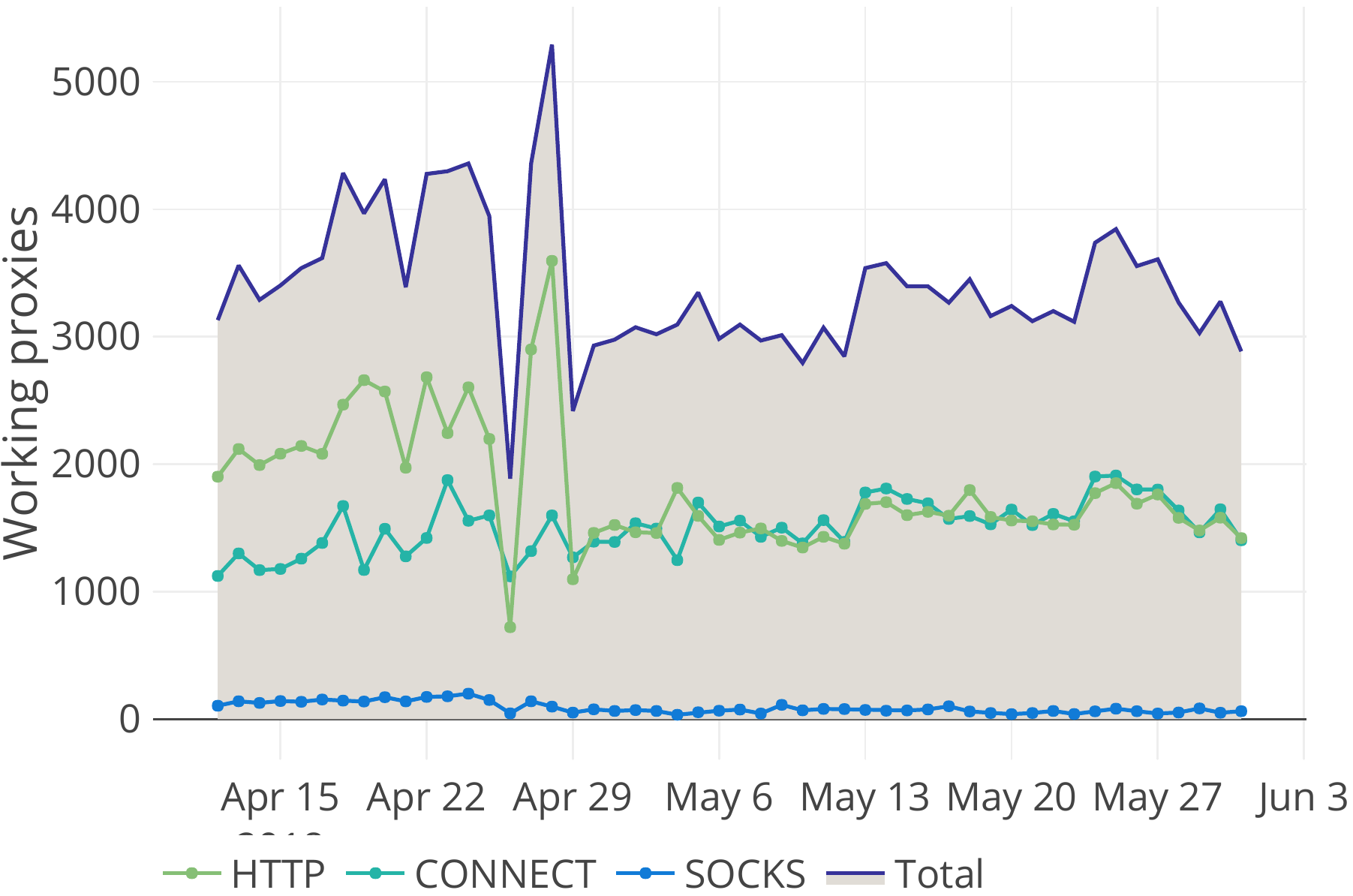}
    \caption{Working proxies, by type, over time.}
    \label{fig:working}
  \end{minipage}
  \hfill
  \begin{minipage}{.32\linewidth}
    \centering
    \includegraphics[width=1.0\linewidth]{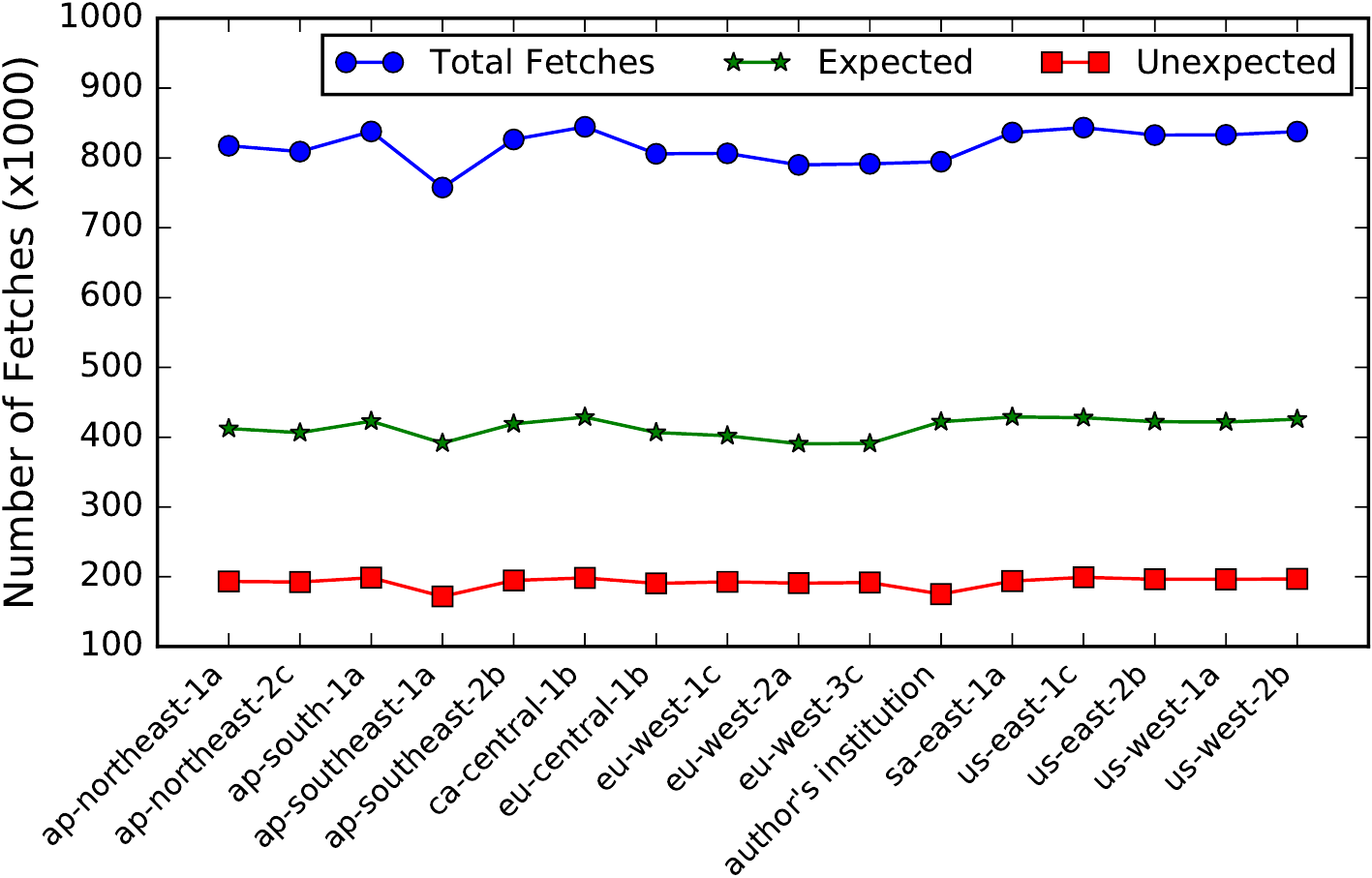}
    \caption{Total fetches and fetches with expected and unexpected content, by client location.}
    \label{fig:regions}
  \end{minipage}
\end{figure*}

The number of unique proxies listed by proxy aggregator sites, over
time, is shown in Figure~\ref{fig:numlisted}.  The median number of
proxies listed on the aggregator sites over all days in the measurement period
is 41,520, with a range of $[38\text{,}843;48\text{,}296]$.  In total,
during the course of our study, we indexed approximately 107,000
unique proxies that were listed on aggregator sites.

{\bf We find that more than 92\% of open proxies that are advertised
  on proxy aggregator sites are offline or otherwise unavailable.}
Figure~\ref{fig:working} plots the number of {\em responsive} proxies
over time for which we were able to establish at least one connection
and retrieve content.  We remark that the Figure includes proxies that
returned unexpected content.  Across our measurement
study, the median daily number of responsive proxies is
\confirmed{3,283}; the medians for HTTP, CONNECT, and SOCKS proxies
are \confirmed{1,613}; \confirmed{1,525}; and \confirmed{74},
respectively.


\begin{table*}[t]
  \begin{minipage}{.74\linewidth}
    
\centering
\begin{scriptsize}
    \begin{tabular}{lr | lr | lr}
    \toprule
    \multicolumn{2}{c|}{{\bf Listed}} & \multicolumn{2}{c|}{{\bf Responsive}} & \multicolumn{2}{c}{{\bf Correct}} \\
    {\bf Country} & {\bf Percent} & {\bf Country} & {\bf Percent} & {\bf Country} & {\bf Percent} \\
    \midrule
China & 18.81\% & Brazil & 17.97\% & Brazil & 19.48\% \\ 
Brazil & 15.55\% & China & 16.75\% & China & 14.74\% \\ 
United States & 15.19\% & United States & 12.12\% & United States & 12.17\% \\ 
Indonesia & 5.95\% & Indonesia & 6.77\% & Indonesia & 7.15\% \\ 
Thailand & 5.14\% & Thailand & 5.73\% & Thailand & 6.06\% \\ 
Russian Federation & 4.60\% & Russian Federation & 5.11\% & Russian Federation & 5.40\% \\ 
Germany & 2.53\% & Singapore & 2.61\% & Singapore & 2.78\% \\ 
Singapore & 2.34\% & Germany & 2.59\% & India & 2.52\% \\ 
India & 2.21\% & India & 2.54\% & Germany & 2.44\% \\ 
Italy & 1.79\% & Canada & 1.93\% & Canada & 1.91\% \\ 
\midrule
All others (count: 145) & 25.69\% & All others (count: 136) & 25.65\% & All others (count: 131) & 25.17\% \\ 

\bottomrule
\end{tabular}
\end{scriptsize}
\caption{Locations of proxies listed on aggregator sites {\em (left)}, capable of accepting proxy connections {\em (center)}, and forwarding correct content at least once {\em (right)}.}
\label{tbl:locations}

  \end{minipage}
  \hfill
  \begin{minipage}{.24\linewidth}
    
\begin{scriptsize}
\begin{tabular}{lr}
    \toprule
    {\bf Software} & {\bf Percent} \\
    \midrule
squid & 87.07\% \\ 
http\_scan\_by & 2.89\% \\ 
1.1 www.santillana.com.mx & 2.29\% \\ 
1.0 PCDN & 1.78\% \\ 
HTTP/1.1 sophos.http.proxy:3128 & 1.74\% \\ 
swproxy & 0.57\% \\ 
Cdn Cache Server & 0.40\% \\ 
MusicEdgeServer & 0.34\% \\ 
1.1 Pxanony & 0.22\% \\ 
1.1 j5k-8 (jaguar/3.0-11) & 0.21\% \\ 
\midrule
All others (count: 120) & 2.50\% \\ 

\bottomrule
\end{tabular}
\end{scriptsize}
\caption{Proxy software.}
\label{tbl:software}

  \end{minipage}
  \vspace{-.2in}
\end{table*}

\begin{table*}[t]
\centering
\begin{scriptsize}
\begin{tabular}{llr | llr | llr}
\toprule
    \multicolumn{3}{c|}{{\bf Listed}} & \multicolumn{3}{c|}{{\bf Responsive}} & \multicolumn{3}{c}{{\bf Correct}} \\
    {\bf ASN} & {\bf Name} & {\bf Percent} & {\bf ASN} & {\bf Name} & {\bf Percent} & {\bf ASN} & {\bf Name} & {\bf Percent} \\ 
\midrule
4134 & No.31,Jin-rong Street & 9.34\% & 14061 & DigitalOcean, LLC & 7.81\% & 14061 & DigitalOcean, LLC & 7.47\% \\ 
14061 & DigitalOcean, LLC & 6.69\% & 4134 & No.31,Jin-rong Street & 7.73\% & 4134 & No.31,Jin-rong Street & 6.65\% \\ 
18881 & TELEFÔNICA BRASIL S.A & 3.55\% & 18881 & TELEFÔNICA BRASIL S.A & 4.10\% & 18881 & TELEFÔNICA BRASIL S.A & 4.52\% \\ 
4837 & CHINA UNICOM China169 Backbone & 2.51\% & 17974 & PT Telekomunikasi Indonesia & 2.63\% & 17974 & PT Telekomunikasi Indonesia & 2.77\% \\ 
17974 & PT Telekomunikasi Indonesia & 2.26\% & 4837 & CHINA UNICOM China169 Backbone & 2.53\% & 4837 & CHINA UNICOM China169 Backbone & 2.40\% \\ 
13335 & Cloudflare Inc & 1.81\% & 45758 & Triple T Internet/Triple T Broadband & 1.97\% & 45758 & Triple T Internet/Triple T Broadband & 2.10\% \\ 
16276 & OVH SAS & 1.73\% & 20473 & Choopa, LLC & 1.66\% & 20473 & Choopa, LLC & 1.82\% \\ 
45758 & Triple T Internet/Triple T Broadband & 1.73\% & 53246 & Cyber Info Provedor de Acesso LTDA ME & 1.62\% & 53246 & Cyber Info Provedor de Acesso LTDA ME & 1.79\% \\ 
20473 & Choopa, LLC & 1.44\% & 16276 & OVH SAS & 1.52\% & 16276 & OVH SAS & 1.56\% \\ 
53246 & Cyber Info Provedor de Acesso LTDA ME & 1.38\% & 31034 & Aruba S.p.A. & 1.33\% & 31034 & Aruba S.p.A. & 1.45\% \\ 
\midrule
--- & All others (count: 3419) & 64.27\% & --- & All others (count: 2961) & 61.17\% & --- & All others (count: 2759) & 59.06\%\\ 

\bottomrule
\end{tabular}
\end{scriptsize}
\caption{Most popular ASNs for proxies listed on aggregator sites {\em (left)}, proxies capable of accepting  connections {\em (center)}, and proxies that forwarded correct content at least once {\em (right)}.}
\label{tbl:asns}
\end{table*}

We use the MaxMind GeoLite2 City and ASN databases~\cite{maxmind} to
resolve each proxy's IP address to a physical location and an
autonomous system (AS).  Tables~\ref{tbl:locations} and~\ref{tbl:asns}
respectively report the most frequent locations and ASes from among
the proxies that are listed, responsive (i.e., respond to proxy
requests), and successfully deliver the expected content at least
once.  {\bf There is surprisingly little geographic and network
  diversity among the proxies.}  Ten countries are responsible for
nearly three-quarters of the world's working proxies, while Brazil
alone is home to nearly 20\% of the proxies that forward expected
content.

Similarly, a handful of ASes are privy to traffic from a
disproportionate amount of the open proxies.  In particular,
U.S.-based DigitalOcean and the Chinese No.~31 Jin-rong Street AS each
host approximately 7\% of the working proxies.  In general, the
distribution of open proxies on the Internet is very skewed, with
roughly 40\% of proxies confined to only 10 ASes.  Moreover, open proxies
are found on only a small fraction of the Internet: although more than
31,000 proxies accepted proxy requests during the course of our
experiment, they resided on just 2,971 (5.8\%) of the Internet's
approximately 51,500 autonomous systems~\cite{caida-as}.

We identified Squid, an open-source cacheing proxy, as the most
frequent proxy software among the relays that (i)~responded to our client
requests and (ii)~inserted a self-identifying {\tt Via} or {\tt X-Via} HTTP
header.  As shown in Table~\ref{tbl:software}, Squid was used by over
85\% of such proxies.  Overall, we identified 130 different
self-reported proxy software systems, although we note that not all
proxies include {\tt Via} or {\tt X-Via} HTTP headers (84.43\% do not)
and that such headers are easily forged and may not actually reflect
the actual software.

\begin{figure*}[h]
  \centering
  \begin{minipage}{.18\linewidth}
    \includegraphics[width=\linewidth]{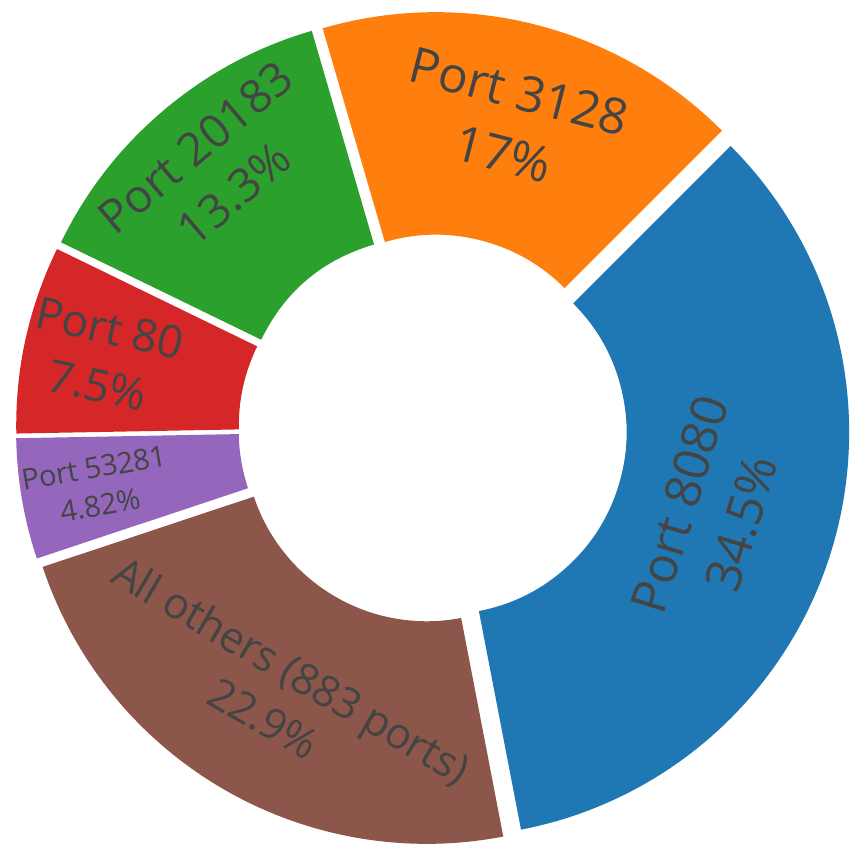}
    \caption{Proxy ports, by popularity.}
    \label{fig:ports}
  \end{minipage}
  \hfill
  \begin{minipage}{.43\linewidth}
    \centering
    \includegraphics[width=1.0\linewidth]{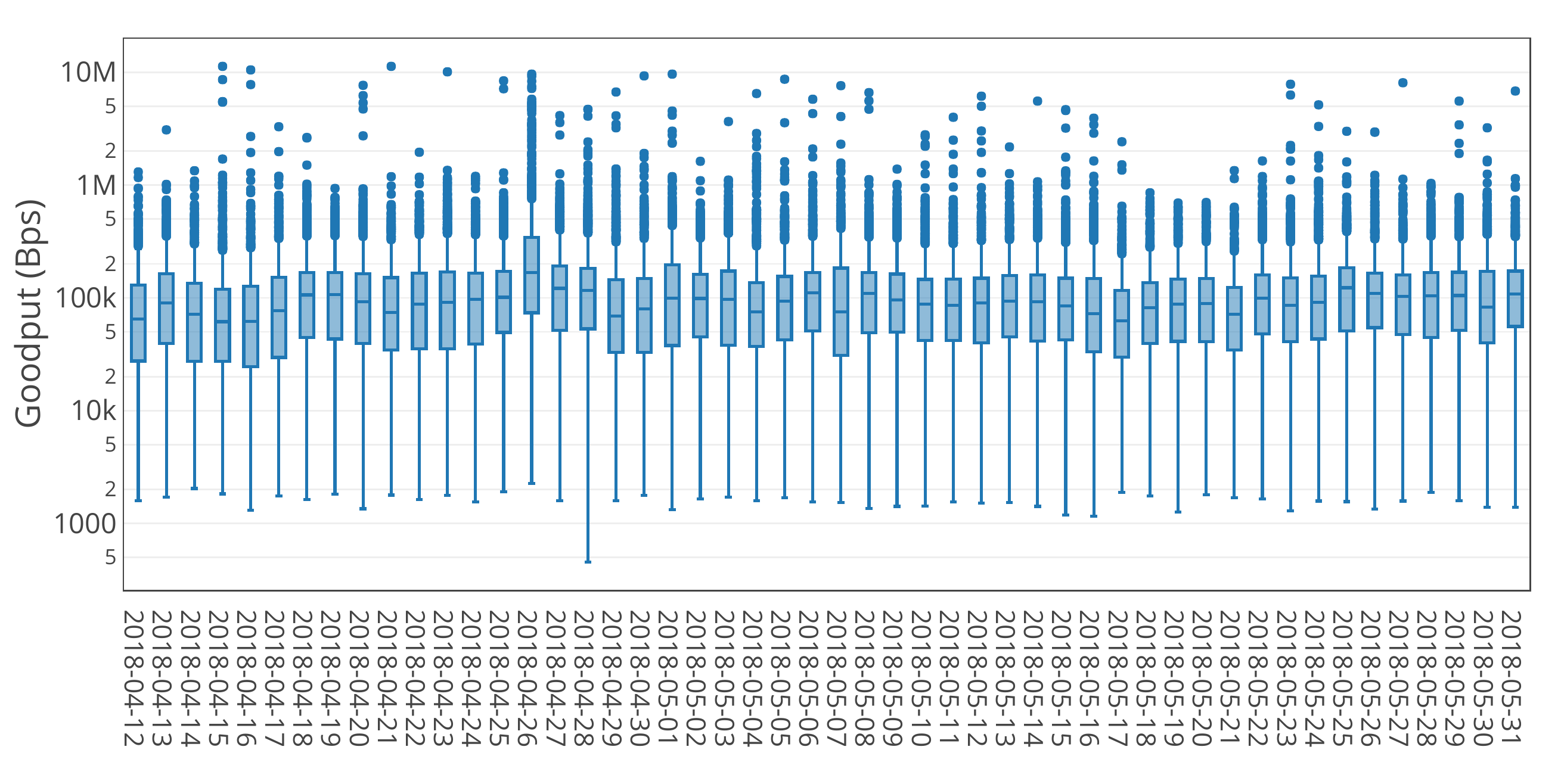}
    \caption{Proxy goodput (log-scale).  Box plots depict the range of
      proxies' average goodput over its daily requests.  The box depicts
      the IQR with the median; the whiskers denote \mbox{1.5 $\times$ IQR}.
      Points beyond the whiskers are outliers.}
    \label{fig:thruput}
  \end{minipage}
  \hfill
  \begin{minipage}{.30\linewidth}
    \centering
    \includegraphics[width=\linewidth]{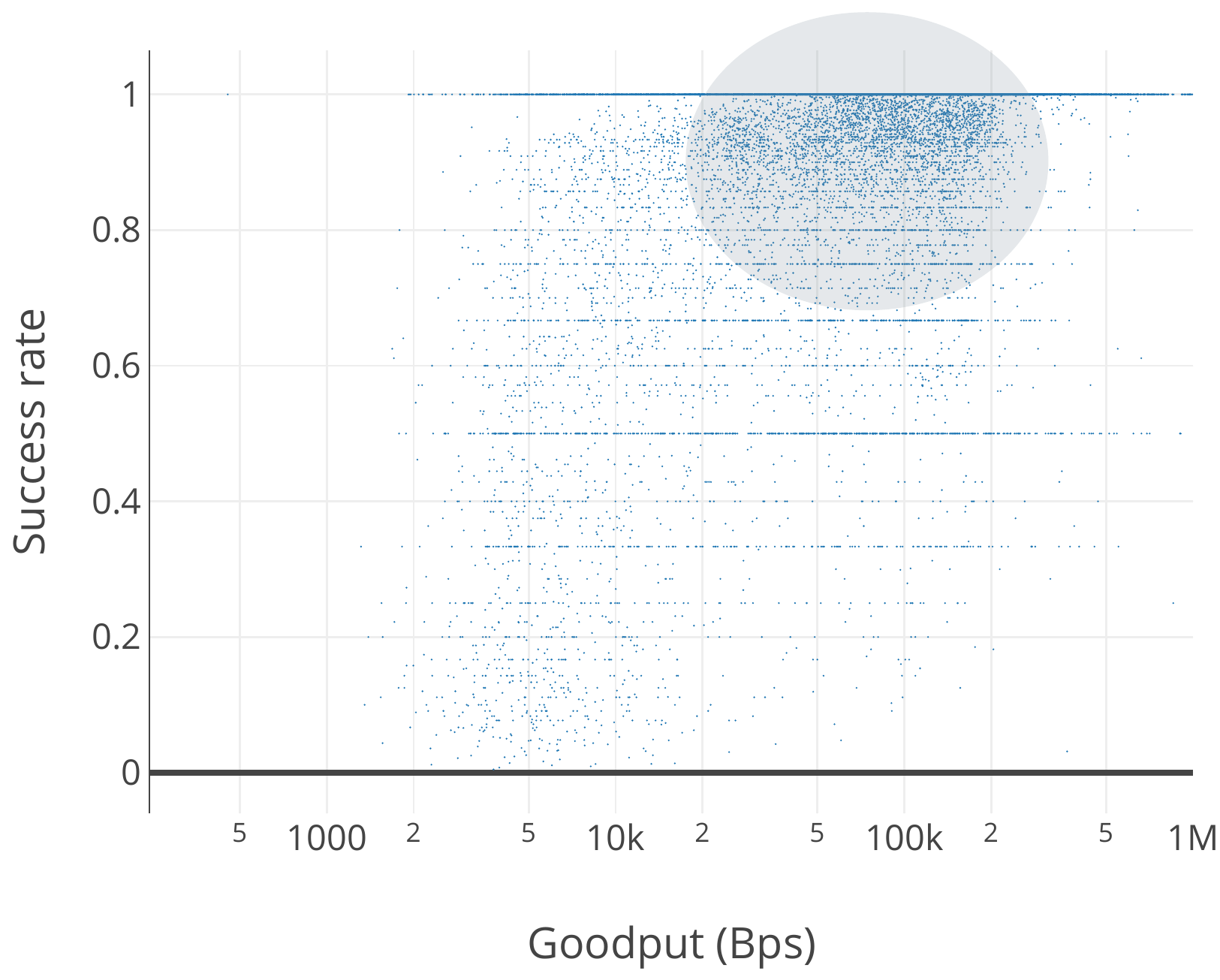}
    \caption{Proxy success rates and average goodput (log-scale).  The
      highlighted oval denotes the densest region of the graph.}
    \label{fig:successrates}
  \end{minipage}
\end{figure*}

As reported in Figure~\ref{fig:ports}, approximately 62\% of working
proxies listened on TCP ports 8080, 3128, or 20183.  Port 8080 is an
alternative port frequently used for web traffic or cacheing servers,
and 3128 is the default port used by Squid.  The popularity of port
20183 is surprising; it is not listed as a standard port by the
Internet Assigned Numbers Authority (IANA).  The standard web port,
port 80, is used by 7\% of open proxies.  The frequent use of ports
8080 and 3128 by open proxies provides a potential means of
discovering additional potential proxies: Censys, which is based on
ZMap~\cite{zmap} Internet scans, reports 7.3M Internet hosts listening
on port 8080; Shodan lists 1.1M hosts with services listening on port
3128~\cite{shodan}.  As we discuss in \S\ref{sec:ethics}, we limit our
study to only the proxies whose addresses are listed publicly by
aggregator sites, and thus for ethical reasons do not probe these
additional 8.4M hosts to discover unlisted proxies.

\subsection{Performance}
\label{sec:performance}

We evaluate the performance of proxies by considering {\em goodput},
computed as the size of a fetched file (we use a 1MiB file) divided by the
time taken to download it through a proxy.  We consider only instances in
which the response reflects the expected content (i.e., the 1MiB
file).  Figure~\ref{fig:thruput} shows the range of proxies' daily
average goodput for requests that yielded the expected content.
Overall, the average goodput for such proxies is
128.5KiBps, with an interquartile range (IQR) of \mbox{[39.5; 160.9]
  KiBps}.

We also consider proxies' utility over time.  A proxy that offers high
goodput but functions only sporadically is not particularly useful.
We define the {\em success rate} to be the fraction of proxy requests
that were successfully completed and yielded the expected content.
Figure~\ref{fig:successrates} plots proxies' success
rates as a function of their average goodput.  We note that,
generally, the proxies that offer the highest goodput (highlighted in
the Figure with the oval) also tend to offer the highest success
rates.  The existence of stratified ``lines'' when the success rate is
$1/3$, $1/2$, and $2/3$ is somewhat surprising: this indicates a
regular periodicity or schedule during which these proxies are
available.  

\subsection{Expected vs.~Unexpected Content}
Proxies do not always return the expected content.  However, not all
unexpected content is malicious.  For example, many tested proxies
return a login page or a page conveying an authentication error,
regardless of the requested URL.  These indicate that the listed open
proxy is either misconfigured, actually a private proxy, or is some
other misconfigured service. Here, we answer the question: to what
extent do listed open proxies return the expected content?

For our analysis, we consider the proxies that have responded to at
least one proxy request with a non-zero byte response and a {\tt 2xx}
HTTP response code that indicates success (i.e., in $[200,299]$).  For
each such proxy, we determine its failure rate, which we define as
$1-\text{success rate}$.  That is, a proxy's failure rate is the
fraction of returned responses that constitute unexpected content.

\begin{figure}[t]
  \centering
  \includegraphics[width=1.0\linewidth]{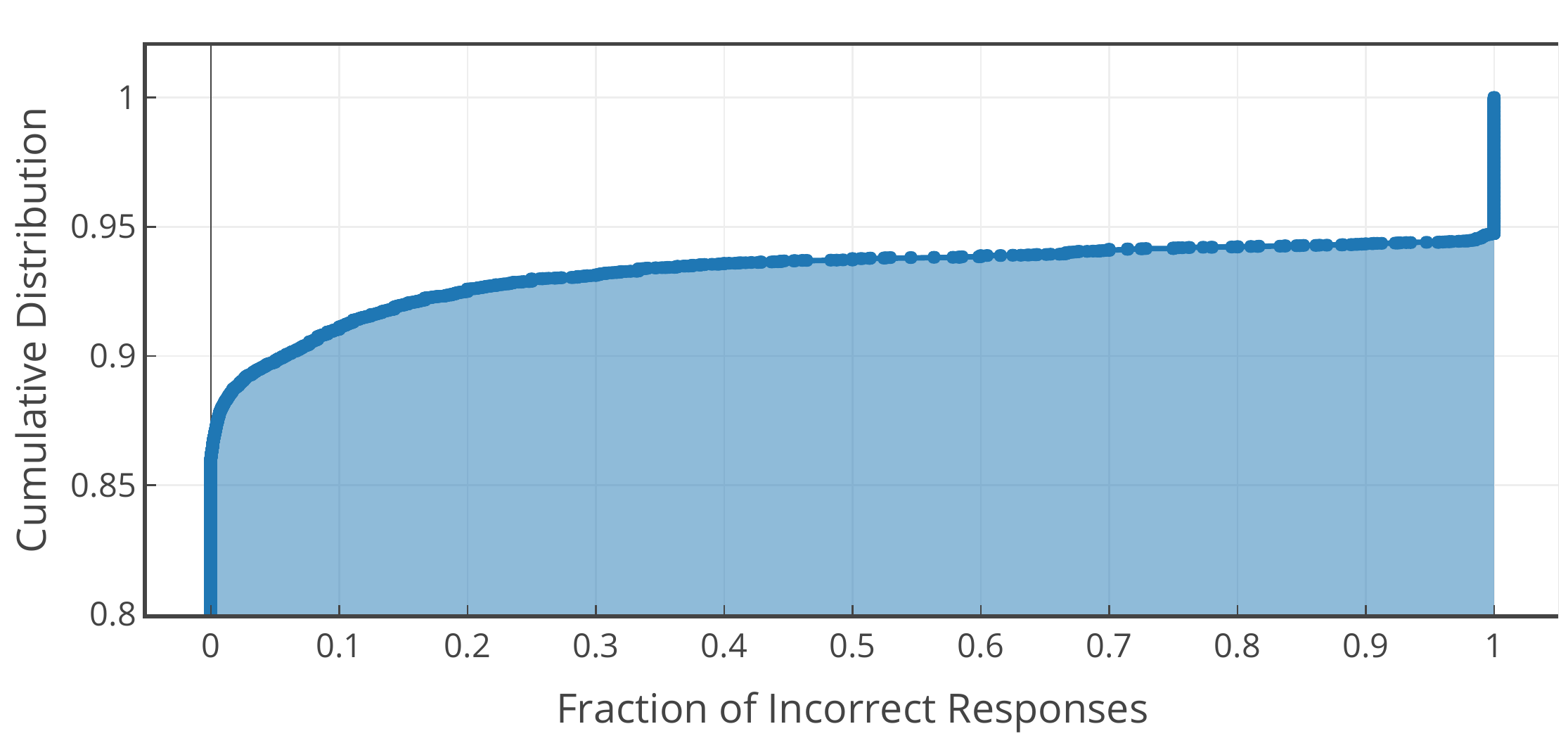}
  \caption{Cumulative distribution of the proxies' failure rate (i.e.,
    the fraction of responses that constituted unexpected content).}
  \label{fig:unexpected}
\end{figure}

Figure~\ref{fig:unexpected} shows the cumulative distribution (y-axis)
of these proxies' failure rates (x-axis).  Of proxies that respond to
proxy requests with {\tt 2xx} HTTP success codes, we find that 92.0\%
consistently deliver the expected content.  {\bf Alarmingly,
  approximately 8\% of the proxies at least sometimes provided
  unexpected content, and 3.6\% of the proxies {\em consistently}
  returned unexpected content}---all with HTTP response codes that
indicate success.  In \S\ref{sec:html} and~\S\ref{sec:files}, we
explore cases in which the content has been purposefully and
maliciously manipulated---for example, to return a trojan .exe file
or to insert spurious ads in retrieved HTML content.
\confirmed{Further, as shown in Figure~\ref{fig:regions}, the
  behavior of proxies---that is, whether they returned expected or unexpected
content---does not significantly vary with the location of the requesting client.}

\subsection{Anonymity}
Open proxies are sometimes used as a simple method of hiding a user's
IP address.  We find that such a strategy is mostly ineffective, {\bf
  with nearly two-thirds of tested proxies exposing the requestor's IP
  address.}

In more detail, we inspect the HTTP {\em request} headers that are
sent by a proxy to the destination webserver.  These include the
headers added by our client toolchain (specifically: {\tt User-Agent} and {\tt
  Accept}) and those inserted by the proxy.  To accomplish this, we
constructed and hosted a simple web application that records HTTP
request headers.  We then examined the request headers that were
produced when we used a client at our local institution 
to access our web application
through each proxy.  Of the proxies that were able to connect to our
web application (13,740), we found that 66.08\% (9,079) inserted at
least one header (most commonly, {\tt X-Forwarded-For}) that contained
the IP address of our client.

\Section{HTML Manipulation \pages{1}}
\label{sec:html}

\begin{figure}[t]
  \centering
  \includegraphics[width=0.85\linewidth]{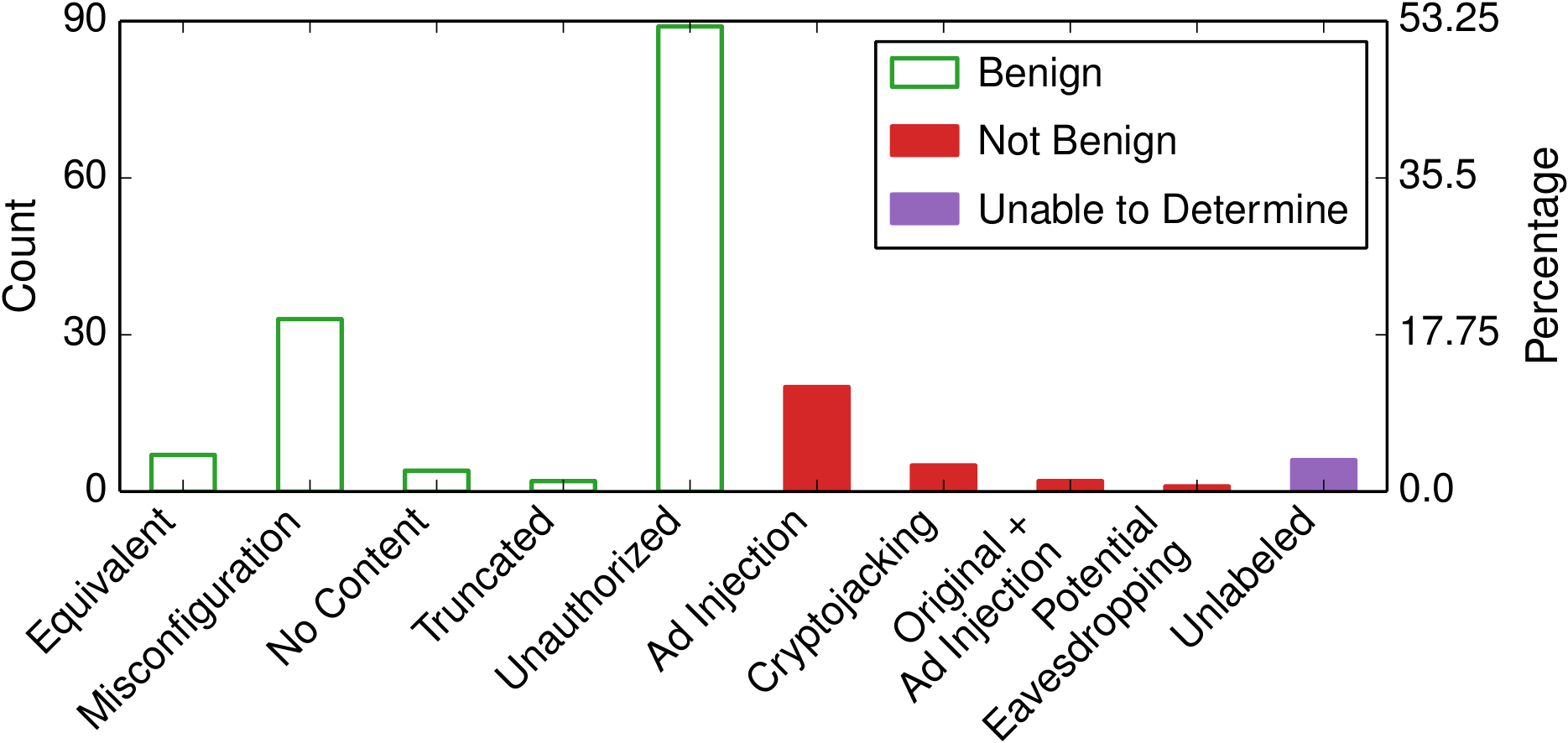}
  \caption{Classification of modified HTML retrieved on \mydate{2018-05-07}.}
  \label{fig:htmlsuspicious}
\end{figure}

We begin our study of malicious behavior among open proxies by
considering the manipulation of fetched HTML content.  In the absence
of end-to-end SSL/TLS (i.e., the use of HTTPS), a malicious proxy can
trivially either modify the web server's response or respond with
arbitrary content of its choosing.

To detect such misbehavior, we fetch an HTML page via the open proxies
each day between \mydate{2018-05-04} and \mydate{2018-05-31}. For
simplicity, we focus on the HTML pages fetched through the client at
our local institution. We observe a median of 1,133 successful
requests (i.e., completed requests with {\tt 2xx} HTTP response code)
per day.

Overall, {\bf we find that 10.5\% of requests for HTML pages
produced {\em un}expected HTML content}.  

As noted above, not all unexpected content necessarily corresponds to
malicious behavior. 
Since the observed fraction of unexpected HTML content is almost the
same for each day of our measurement study, we perform an in-depth
analysis for a random day, \mydate{2018-05-07}, to determine whether
the HTML manipulation could be categorized as malicious.  On that
date, we observe requests through 1,259 proxies, of
which 169 (about 13.4\%) return unexpected HTML content.  (We discard
requests that did not yield {\tt 2xx} success HTTP response codes.)

We manually analyze these 169 \textit{modified} HTML files. Our
methodology is as follows. First, we inspect each file and divide
them into two categories: benign (non-malicious) files and suspicious files.  The
latter category represents HTML files that contain Javascript code
(either directly in the file or through the inclusion of an external
Javascript file).

We then collectively inspect each of these benign and suspicious
files and classify them into five benign classes and four non-benign
classes. This inspection process involves manually examining the
files, rendering them in a browser on a virtual machine, and
potentially visiting Javascript URLs that were inserted.  The five
classes that we posit do {\em not} indicate maliciousness
are as follows:
\begin{itemize*}
\item {\em Equivalent:} the fetched HTML renders equivalently in
  a web browser to the expected content.  Oddly, we found instances in
  which HTML tag attributes were inconsequentially reordered; for
  instance, tags such as
  \html{<IMG SRC="A" ALT="B">} are rewritten as \html{<IMG ALT="B" SRC="A">}.

\item {\em Misconfiguration:} the retrieved content constituted
  error pages, often displaying ``not accessible'' messages.

\item {\em No Content:} the page contained no content.

\item {\em Truncated:} the retrieved page was a truncated
  version of the expected content.

\item {\em Unauthorized:} the pages showed
error messages such as ``invalid user'' or ``access denied''.
\end{itemize*}

The identified classes of proxy misbehavior are:
\begin{itemize*}
\item {\em Ad Injection:} the proxy replaced HTML
content with Javascript that rendered extraneous advertisements.
\item {\em Original + Ad Injection:} the returned HTML contained the
  expected content, but also included a single ad injection.
\item {\em Cryptojacking:} the returned HTML include Javascript code
  that would cause the user's browser to mine cryptocurrencies on
  behalf of the proxy.
\item {\em Potential Eavesdropping:} the retrieved page contained
  Javascript that caused the user's browser to visit pages from its
  history, potentially revealing these pages to the proxy if they were
  previously visited without the use of the proxy.
\end{itemize*}

The classifications of the modified HTML files with their counts and
percentages are shown in Figure~\ref{fig:htmlsuspicious}.  We
were unable to classify six HTML responses, described in the Figure as
``Unlabeled''.

Overall, we find that approximately 80\% of the unexpected HTML responses are benign. Approximately 72\% of the pages
contained errors, likely due to private access proxies or due to
misconfigured proxy software.


 {\bf We find that \confirmed{16.6\%} of
  unexpected HTML responses (and 2.2\% of overall proxy requests for
  HTML content) correspond to malicious activity,
  including ad injection, cryptojacking, and eavesdropping.}  Among the
malicious activity, most prominent (13\%) is ad injection.


\begin{figure}[t]
  \centering
{
\scriptsize
\begin{Verbatim}[frame=single,commandchars=\\\{\}]
\textcolor{blue}{<script} \textcolor{green}{src=}\textcolor{orange}{"https://www.hashing.win/scripts/min.js"}\textcolor{blue}{>}
\textcolor{blue}{</script>}
\textcolor{blue}{<script>}
    \textcolor{blue}{var} miner = \textcolor{blue}{new} Client.Anonymous(
        \textcolor{black}{'\censor{<64 characters long alphanumeric string>}'},
        \{ throttle: 0.1 \}
        );
    miner.start();
\textcolor{blue}{</script>}
\end{Verbatim}
}
\caption{Injected HTML code with link to remote cryptojacking
  Javascript. The user identifier has been redacted.}
\label{fig:js-code-snippet}
\end{figure}

Perhaps most interestingly, we find that about 3\% of the files include Javascript that performs
cryptojacking---that is, the unauthorized use of the proxy user's
processor to mine cryptocurrencies in the background. Upon further
inspection, we determined that each such instance uses the same
injected Javascript code, shown in
Figure~\ref{fig:js-code-snippet}.

The referenced \html{min.js} script is obfuscated Javascript.  After
decoding it, we determined that it is similar to
Coinhive's~\cite{coinhive} Monero\footnote{Monero is advertised as a
  ``secure, private, and untraceable'' cryptocurrency~\cite{monero}.}
cryptocurrency~\cite{monero} mining Javascript. The 64 character long
argument to the Client constructor serves as the identifier for the
user to be paid in the original Coinhive setting. (We redact this
argument in Figure~\ref{fig:js-code-snippet} because it potentially
identifies a criminal actor.)  Finally, we note that all Coinhive
endpoints described in the copied Coinhive script are replaced with
other domains, indicating that whoever is running the infrastructure
to collect the mining results is not using Coinhive's service.

\Section{File Manipulation \pages{1.5}}
\label{sec:files}

\def\totalProxies{21,385\xspace}

%

\begin{table*}[t]
\small
\centering

\begin{tabular}{l|lllllllll}
\toprule
\multicolumn{1}{c|}{\textbf{\begin{tabular}[c]{@{}c@{}}\\Requested\\ File Type\end{tabular}}} & \multicolumn{9}{c}{\textbf{\begin{tabular}[c]{@{}c@{}}Return File Type (VirusTotal)\end{tabular}}}                                                                \\
                                                                                             & \textbf{EXE} & \textbf{Flash} & \textbf{JAR} & \textbf{ZIP} & \textbf{HTML} & \textbf{ISO} & \textbf{GZIP} & \textbf{XML}  & \textbf{Unknown} \\
\midrule
EXE                                                                                          & 6614/6713          & -              & -            & -            & 410/423       & -            & 0/39          & 0/34          & 12/234           \\
Flash                                                                                        & -                  & 0/1546         & -            & -            & 352/362       & -            & 0/3           & 0/34          & 0/32             \\
JAR                                                                                          & -                  & -              & 0/1075       & 2/400        & 338/351       & -            & 0/2           & 0/34          & 0/55             \\
ZIP                                                                                          & -                  & -              & -            & 12/4162      & 380/395       & -            & 0/18          & 0/34          & 0/176            \\
TEXT                                                                                         & -                  & -              & -            & -            & 436/446       & 545/545      & 0/2           & 0/34          & 23/10910         \\
BAT                                                                                          & -                  & -              & -            & -            & 556/574       & -            & 0/98          & 0/34          & 2/719            \\
\bottomrule
\end{tabular}
\caption{Requested file types and various response file types determined by VirusTotal for unexpected responses. Each entry for a given request and response file type shows the fraction of malicious responses.}
\label{tbl:url-response-type}
\end{table*}

\newcolumntype{L}[1]{>{\raggedright\arraybackslash}p{#1}}
\newcolumntype{C}[1]{>{\centering\arraybackslash}p{#1}}
\newcolumntype{R}[1]{>{\raggedleft\arraybackslash}p{#1}}

\begin{table*}[h]
\centering
\begin{small}
\begin{tabular}{L{25mm}|C{10mm}C{7mm}C{8mm}C{8mm}C{8mm}C{8mm}C{5mm}C{8mm}}

\toprule
    \textbf{\begin{tabular}[c]{@{}l@{}} No.~of antivirus\\systems flagging\\as malicious\\\end{tabular}} & {\bf 0} & {\bf 1} & {\bf 2} & {\bf 3} & {\bf 4} & {\bf 5} & {\bf 6} & {\bf 18} \\
\midrule
              {\bf No.~of Files} & 19,802   &   587   &  1,274   &  1,508   &  4,232   &  1,513  &  23   &   545 \\
 &  (67.16\%) & (1.99\%) & (4.32\%) & (5.11\%) & (14.35\%) & (5.13\%) & (0.08\%) & (1.85\%) \\
\bottomrule
\end{tabular}
\end{small}
\caption{The number of files and percentage (relative to all unexpected content files) {\em (bottom row)} flagged as malicious by varying number of antivirus systems {\em (top row)} used by VirusTotal.}
\label{tbl:malicious-files-count}
\end{table*}


\begin{figure}[t]
  \centering
  \includegraphics[width=1.0\linewidth]{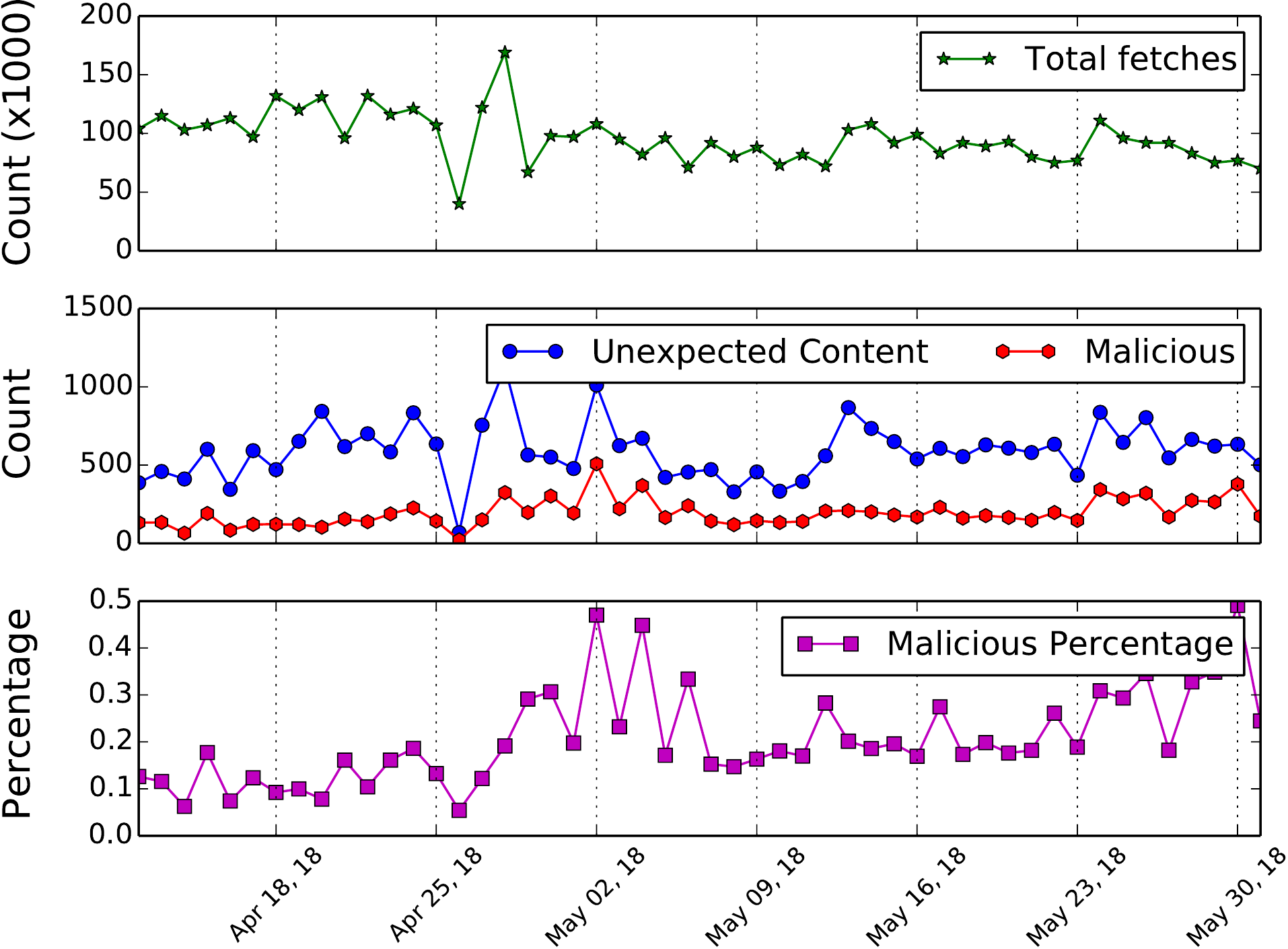}
  \caption{Per day total requests (\textit{top}), unexpected responses and malicious responses (\textit{middle}),
  and percentage of malicious responses relative to total responses (\textit{bottom}).}
  \label{fig:file-manipulation-day-wise}
\end{figure}

Over the duration of our 50 day experiment, we made
more than 4.8M 
successful requests
through 21,385 proxies
 to
fetch a variety of non-HTML files (specifically, Windows .exe, Java
.jar, Flash, .zip, and Windows .bat files).  As
before, we define a successful request to be one in which the proxy
returns a {\tt 2xx} HTTP response code and a non-zero content size.
Furthermore, we exclude HTML, plaintext, and PHP responses (as determined by filemagic) from our
analysis to avoid uninteresting instances of error pages or
unauthorized access pages (see \S\ref{sec:html}). 

Figure~\ref{fig:file-manipulation-day-wise} (\textit{top}) shows the frequency of total successful
requests made per day, averaging approximately 97K
requests per day throughout the measurement
period. Overall, 29,484 (\confirmed{0.61\%}) of such requests
(made by \confirmed{6.76\%}  of the proxies), constitute unexpected
content. 

To check whether unexpected content is malicious, we submitted all
\confirmed{29,484} unexpected responses to VirusTotal~\cite{virustotal} for scanning. VirusTotal
scans uploaded files using multiple antivirus tools and returns
detailed analysis results, including the uploaded files determined
file types.
Table~\ref{tbl:url-response-type} provides a summary of our findings. Each entry for
a given request and response type (determined by VirusTotal) indicates the fraction of responses
with unexpected content that were flagged as malicious by VirusTotal. Note that VirusTotal
(correctly) determined various responses to be HTML that were wrongly classified as non-HTML by
filemagic in our initial pass of filtering out HTML pages. 


To establish a baseline, we verified that all of the {\em expected
  responses} are not classified by VirusTotal as being malicious.
That is, no single antivirus tool used by VirusTool marks the
expected content as being malicious, as expected.

On the
contrary, VirusTotal flags \confirmed{32.84\% (9,682/29,484)} of the unexpected responses as
malicious; that is, at least one of the antivirus systems used by
VirusTotal flags the content as malicious.
Figure~\ref{fig:file-manipulation-day-wise}
(\textit{middle}) shows the number of unexpected and malicious responses per day. The percentage of
malicious responses (relative to total success successful fetches per day) remained fairly consistent throughout
our study (Figure~\ref{fig:file-manipulation-day-wise}
(\textit{bottom})).  {\bf Across our experiment period, we find that,
  on average, \confirmed{0.2\%} of daily proxy responses are classified as malicious by
  at least one antivirus system used by VirusTotal.}

Table~\ref{tbl:malicious-files-count} shows the number of unexpected responses that are flagged as
malicious, broken down by the number of antivirus tools used by
VirusTotal that classified it as malicious.
At the extreme, for
example, we find 545 retrieved files that are considered malicious by
18 different antivirus systems.


\subsection{Detailed Findings}
\label{sec:details}
In what follows, we discuss in more detail our
findings of malicious proxy activity, organized by file type.




\Paragraph{Windows Executables (.exe)}
Almost all (\confirmed{98.5\%}) 
of unexpected responses
for .exe files are classified as malicious.  {\bf We find that \confirmed{1.93\% (413/\totalProxies)} of the proxies
maliciously modified the .exe file at least once during our measurement period.} The
infections include malware from the Expiro malware family that can be used to steal personal information and
provide remote access to the attacker~\cite{expiro}, and flavors of malware from the Crypt and Artemis
trojan families~\cite{crypt-trojan, artemis-trojan}. Table~\ref{tbl:win32-top-infections} in
Appendix~\ref{app:file-manipulation} lists the top 10 reported infections for .exe files.

\Paragraph{Flash and .jar Files} VirusTotal did not flag any of the
modified Flash or Java .jar files as malicious.  We are unsure of
whether this indicates (i)~benign (but unexplained) instances in which
proxies rewrite these files, or (ii)~a limitation of the scanners
used by VirusTotal.


\Paragraph{ZIP Files} For ZIP file responses with unexpected content, only a single antivirus system
(McAfee-GW-Edition, v2017.2786) flagged 0.28\% (14/4562) of them as malicious. VirusTotal  did not
provide any details about the infection. 


\Paragraph{HTML Files} 
We
 received HTML responses from proxies (as determined by VirusTotal) irrespective of the
requested content type. 
Roughly  0.2\% (44/\totalProxies) of the proxies returned unexpected content
at least once, with 43 of them returning malicious content as well. Almost 97\% (2,472/2,551) of these
HTML responses were labeled as malicious. Table~\ref{tbl:unexpected-html-infections}
in Appendix~\ref{app:file-manipulation}  shows the reported infections and the number of HTML
responses with malicious code. 
Upon further manual examination, we found that all of the 2,472 malicious responses detected by VirusTotal contain the same
Monero cryptocurrency~\cite{monero} mining Javascript shown in Figure~\ref{fig:js-code-snippet}.



\Paragraph{ISO Files} Surprisingly, on 545 occasions, we  received ISO files when requesting a
1MiB text file.  All of the 545 ISO responses were exactly
1MiB in size, but were flagged as malicious by VirusTotal. Table~\ref{tbl:unexpected-iso-infections}
in Appendix~\ref{app:file-manipulation} shows the various infections reported by VirusTotal.
We note that all 545 files were infected with the Vittalia Trojan, a rootkit for Windows.
The content of 520 of the 545 ISO images
 was identical while the remaining 25 responses were identical to one another. Although the ISO files were clearly malicious, only 0.04\%
(9/\totalProxies) of the proxies returned a malicious ISO response at least once.





\Paragraph{Shell Scripts}
We fetched shell scripts to determine if malicious proxies would modify (or replace) the scripts in transit.
We find that \confirmed{22.75\% (211,288/928,431)} of the requests for shell scripts result in responses with
unexpected content.  To hone-in on malicious activity, we discarded responses whose MIME-type was
not ``text/x-shellscript''; we found \confirmed{1,020} instances in which the responses were unexpected but were
determined to be shell scripts.  Oddly, all 1,020 instances of unexpected content
correspond to just four unique responses, summarized in Table~\ref{tbl:shell-script-results} in Appendix~\ref{app:file-manipulation}.
All of these modifications appear to be non-malicious (strangely, one frequent modification inserts the text ``Pop HerePop HerePop HerePop HerePop
HerePop HerePop HerePop Here'' and little else) and are most
likely due to misconfiguration.   Overall, we
did not find any
evidence of malicious shell script manipulation during our measurements.

\begin{table}[t]
\centering
\begin{scriptsize}
\begin{tabular}{ccc}
\toprule
  {\bf No.~Malicious Proxies} & {\bf AS Number} & {\bf AS Name}\\
\midrule

72 & 4134 & No.31,Jin-rong Street \\
45 & 14061 & DigitalOcean, LLC \\
25 & 4837 & CHINA UNICOM China169 Backbone \\
14 & 17974 & PT Telekomunikasi Indonesia \\
9 & 56046 & China Mobile communications corporation \\

\bottomrule
\end{tabular}
\end{scriptsize}
\caption{The five ASes with the most number of malicious proxies.}
\label{tbl:malicious-proxy-as}
\end{table}

\subsection{Network Diversity and Consistency of Malicious Proxies}
For the responses that we deemed malicious,
we also looked at the distribution of responsible proxies.
Table~\ref{tbl:malicious-proxy-as} shows the top five
ASes with the most number of proxies that performed malicious manipulations.

\begin{figure}[t]
  \centering
  \includegraphics[width=.9\linewidth]{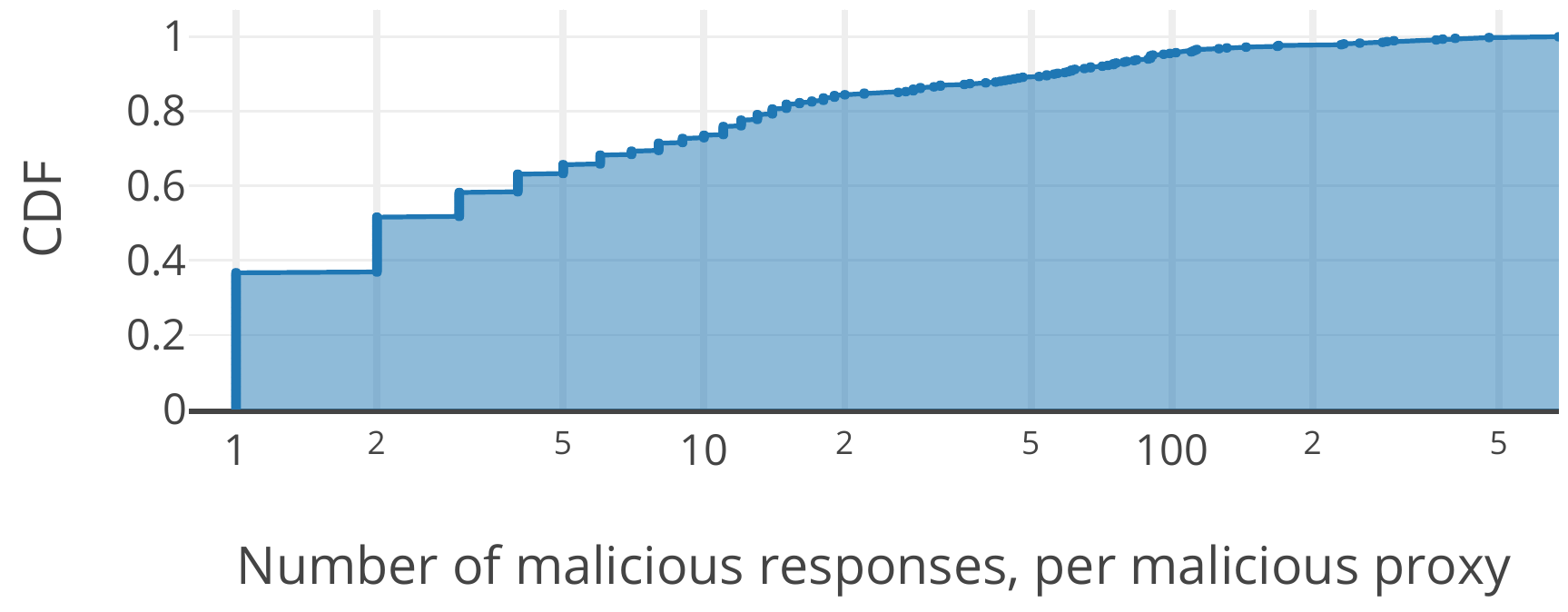}
  \caption{Consistency of malicious proxy behavior.  }
  \label{fig:consistency}
\end{figure}

We also looked at the daily behavior of the top three malicious
proxies over the duration of our measurement, as determined by the
number of malicious responses returned.  The most malicious
proxy 
returned malicious content 100\% of the time. The second and third
most malicious proxies returned malicious responses 97.7\% and 87.9\%
of the time, respectively, when they were reachable.  This trend is
apparent in Figure~\ref{fig:consistency}, which plots (in log-scale)
the cumulative distribution of the number of times that malicious
proxies (defined as a proxy that ever behaves maliciously) exhibit
misbehavior.  Here, we see that half of proxies that return malicious content do so at least twice.
More than 25\% of malicious proxies return at least 10 files with malicious content, while the top 10\% of malicious proxies return 56 or more
malicious files.

Surprisingly, none of the 469 discovered proxies that return malicious
content (\S\ref{sec:details}) are listed on the service run by
Tsirantonakis et al.~\cite{tsirantonakis2018large} that reports
misbehaving proxies.  This suggests that correctly identifying
misbehaving proxies is very challenging, since proxy misbehavior may
be transient and may take different forms.


\Section{SSL/TLS Analysis \pages{1}}
\label{sec:ssl}


We find that 70\% (14,607/20,893) of the
proxies that fetch expected content at least once
allow TLS traffic to pass through them.
We emphasize that supporting HTTPS incurs no overhead on behalf
of the proxy, since the proxy is not involved in the (end-to-end)
cryptography and is merely transporting the ciphertext. It is worth
noting that as of early June 2018, more than 70\% of loaded web
pages were retrieved using HTTPS~\cite{letsencrypt-stats}. The lack of
universal HTTPS support
among the open proxies effectively forces some users to downgrade their
security, bucking the trend of moving towards a more secure web.

We  consider misbehavior among two dimensions: attempts at decreasing
the security of the communication through TLS MitM or downgrade attacks,
and alteration of the content.

\Paragraph{SSL/TLS Stripping} We first examine whether any of the proxies
rewrite HTML \html{<A>} link tags to downgrade the transport from HTTPS to HTTP. For example,
a malicious proxy could attempt to increase its ability to eavesdrop and modify
unencrypted communication by replacing all links to \url{https://example.com} with
\url{http://example.com} on all webpages retrieved over HTTP. To perform this
test, we fetched via each proxy a HTML page hosted on a web server at our institution
over HTTP. This HTML page contained six HTTPS links. We did not find evidence of proxies stripping SSL by replacing the included links.

\Paragraph{SSL Certificate Manipulation} In this experiment, we search for instances
of SSL/TLS certificate manipulation.  While interfering with an HTTPS connection would
cause browser warning messages, numerous studies have shown that even security-conscious
users regularly ignore such warnings~\cite{cryingwolf, warningland}. 

\begin{figure}[t]
  \centering
  \includegraphics[width=0.9\linewidth]{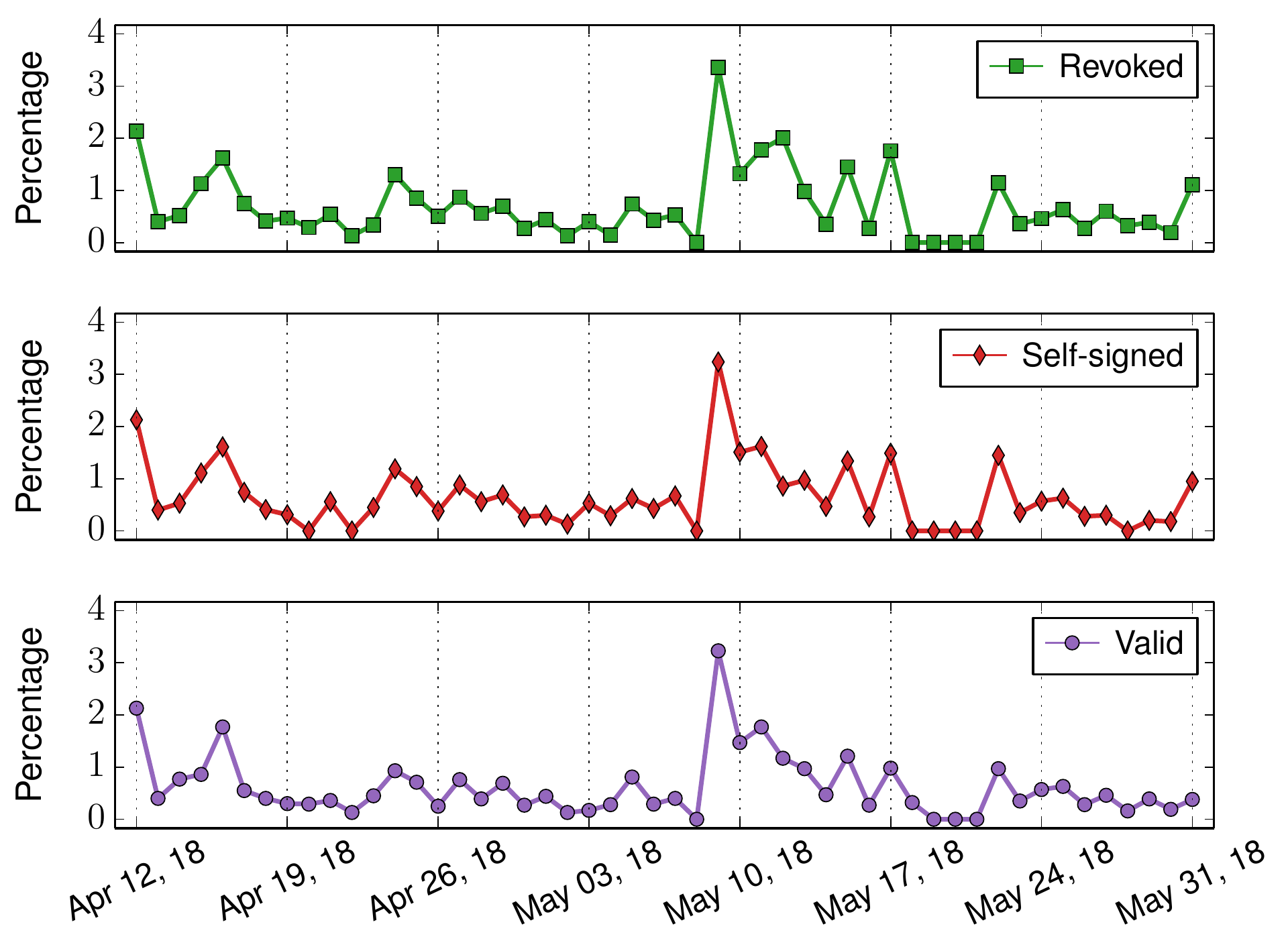}
  \caption{Percentage of proxies that return expected content but perform TLS/SSL MitM, for websites that use revoked (\textit{top}), self-signed (\textit{middle}), and valid and unexpired certificates (\textit{bottom}).}
  \label{fig:sslcertmanipulation}
\end{figure}

We test the proxies against two categories of domains: one with a valid
and verifiable SSL certificate hosted on a web server at our institution;
and domains with incorrect or invalid SSL certificates (\url{https://revoked.badssl.com},
\url{https://self-signed.badssl.com/}). We include the latter category since
we posit that a smart attacker might perform SSL MitM only in instances in
which the connection would otherwise use revoked or self-signed certificates;
in such cases, the browser would issue a warning even in the absence of the
proxy's manipulation.

To detect SSL/TLS certificate manipulation, we fetch the three domains (listed above)
via the open proxies each day between \mydate{2018-04-12} and \mydate{2018-05-31}. For
simplicity, we focus on the pages fetched through the client at our local institution. 

Overall, {\bf we find that 1.06\% (102/9,625) of
proxies that support HTTPS perform TLS/SSL MitM by inserting a modified certificate.}
To determine whether any of the proxies performing SSL/TLS  MitM might
belong to a known botnet, we searched the SSL
Fingerprint Blacklist and Dyre SSL Fingerprint Blacklist~\cite{sslbl}
for the modified
certificates' fingerprints.
We did not find any blacklisted certificates.

We next consider proxies that fetch the {\em expected content} but modify the SSL/TLS certificate.
Such behavior suggests that the proxy is eavesdropping on HTTPS connections. The percentage of
such eavesdropping proxies, per day, for the different categories of
certificates is plotted in
Figure~\ref{fig:sslcertmanipulation}. Overall, 0.85\% (82/9,607) of the proxies that return expected responses appear
to be eavesdropping. We did not find any evidence of proxies selectively targeting incorrect or invalid
SSL certificates.

Finally, we analyze the modified TLS certificates inserted by the  eavesdropping proxies when the requested site is
\url{https://revoked.badssl.com} or
\url{https://self-signed.badssl.com/} (i.e., when the genuine certificate is revoked or self-signed, respectively).
We find that there were 435 modified certificates from 21 unique issuers.
The issuer common name (CN) strongly suggests that 19\% (4/21) of the issuers were schools. We posit that
the proxies that modified these certificates were operated by schools and were incorrectly configured to serve
requests from any network location. Interestingly, all of the 
certificates
inserted by these school proxies had the expected subject common name and were valid (but not normally verifiable,
since the school is not a root CA).  This leads to an interesting result: if the schools pre-installed root CA certificates
on students' or employees' computers, then they are significantly degrading the security of their users.  That is, they are masking
the fact that requested webpages have expired or revoked certificates by replacing these invalid certificates with ones that would be verified.
This is similar to the effects observed by Durumeric et al.~\cite{durumeric2017security} in their examination of enterprise middleware boxes, which also degraded security by replacing invalid certificates with valid ones signed by the enterprise's CA.

Performing TLS MitM also allows a malicious proxy to modify page content. To detect such behavior, using the same approach as described in \S\ref{sec:html}, we analyzed HTML pages fetched over TLS via the open proxies. We did not find any malicious activity.

\Section{Comparison With Tor}
\label{sec:tor}

Tor~\cite{tor} provides anonymous TCP communication by routing
user traffic through multiple relays (typically three) using
layered encryption. The first relay in the path (or circuit) is
the guard relay, and the final relay through which traffic exits
is the exit relay. The original data transmitted is visible only
at the exit relays. Therefore, unless e2e encryption is
used, data can be eavesdropped by malicious or compromised exits.
Prior research studies have found evidence of malicious behavior
such as interception of credentials, etc., by a small fraction of
Tor exit relays, especially when the traffic was not e2e encrypted~\cite{chakravarty2011detecting, tor-usage,SpoiledOnions}.
We now compare the level of manipulation as observed using open
proxies to that when content is fetched over Tor.

We modify Exitmap~\cite{SpoiledOnions}, a fast scanner that
fetches files through all Tor exit relays. To maintain consistency
with our earlier experiments, we fetched the same set of
files (e.g., HTML, .exe, etc.) as described in \S\ref{sec:html} and
\S\ref{sec:files} over HTTP, and accessed the same HTTPS URLs as
described
in \S\ref{sec:ssl}.
We fetched these files each day through every available
Tor exit relay between \mydate{2018-05-06}
and \mydate{2018-05-31}, during which the median number of available exit relays
was 722.

Approximately 13.8\% of connections and 1.8\% of fetches timed out when
using Exitmap.  This is unsurprising since Exitmap does not perform
the same (and necessary) bandwidth-weighted relay selection  as  the standard Tor client~\cite{SpoiledOnions,tor}.

Over our 26 day Tor experiment, {\bf we
found no instance in which a Tor exit relay manipulated either file contents or SSL/TLS certificates.}
Comparing our results to \S\ref{sec:html}-\ref{sec:ssl}, this strongly
suggests that Tor is a more trustworthy network for
retrieving forwarded content.  However, since Tor exits may still passively
eavesdrop (which we would not detect), we concur with the conventional wisdom
that e2e encryption (i.e., HTTPS) is appropriate when using Tor.

\begin{figure}[t]
  \centering
  \includegraphics[width=1.0\linewidth]{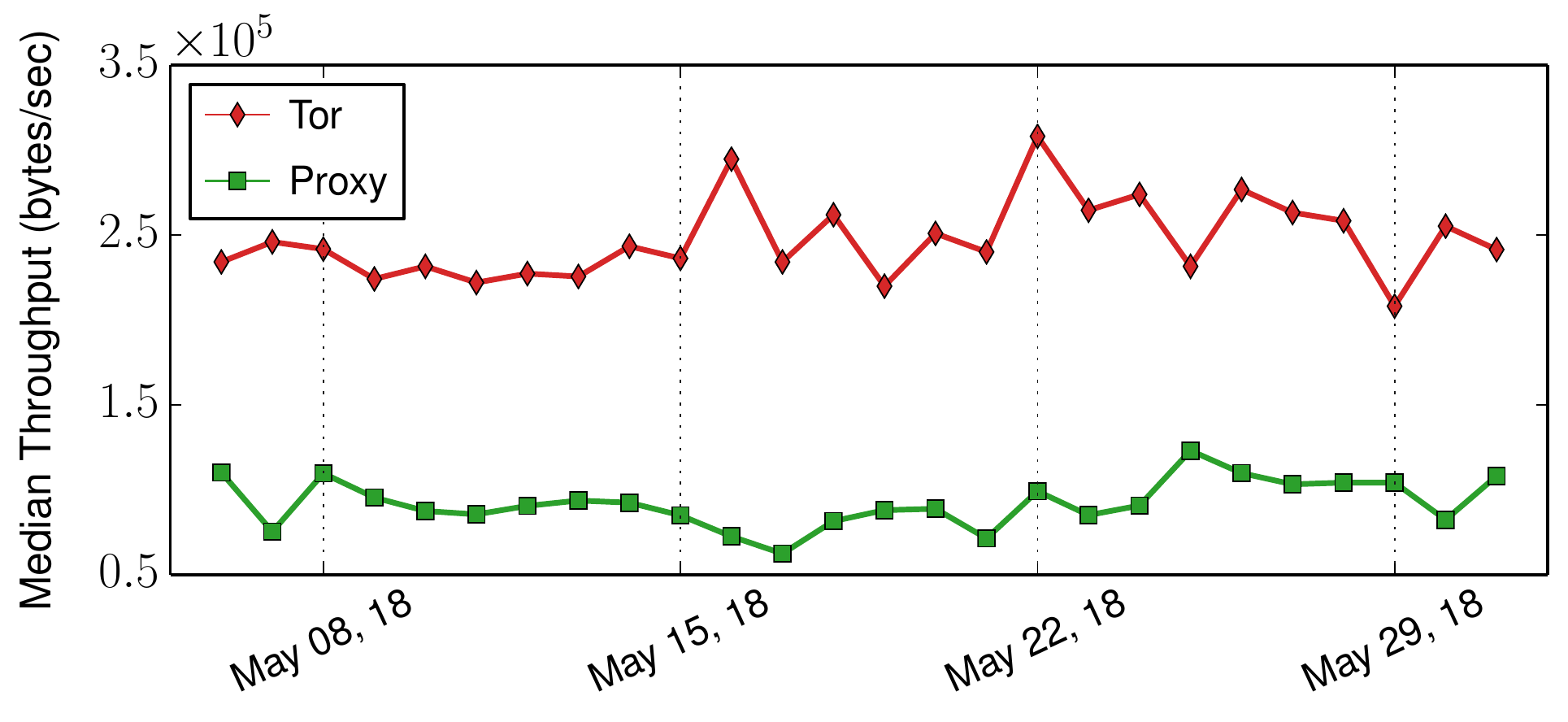}
  \caption{Tor vs. open proxy median throughput per day while fetching a static file of size 1~MiB.}
  \label{fig:median-thruput}
\end{figure}

We also compare the performance of Tor to that of the open proxies. We compute the open
proxies throughput as described in \S\ref{sec:performance}. For Tor,
we rely on data from  the Tor
Metrics Portal~\cite{tormetrics}; specifically, we use the median time taken per day to download a static
file of size 1~MiB and derive Tor's median throughput between \mydate{2018-05-06}
and \mydate{2018-05-31}. Figure~\ref{fig:median-thruput} shows the
 Tor and open
proxies' median  throughput per day. We find that the Tor median
throughput is roughly twice the open proxy throughput. This suggests that Tor performs relatively faster than the open proxies.

\Section{Ethical considerations}  
\label{sec:ethics}

We designed and conducted our measurements to minimize risk.  In
particular, we believe our study's design and implementation meet the
criteria described in the Menlo Report~\cite{menlo-report}, an ethical
framework for conducting network measurements that has been widely
adopted by the networking and computer security communities.  The
Menlo Report describes four principles of ethical research. 

We achieve the principle of {\em Respect for Persons} by avoiding the collection of
data belonging to individual users.  In our experiments, we use open
proxies largely as they are intended: we issue standard well-formed proxy
requests and request only benign (non-malicious) traffic from
non-controversial websites.  We record only our own traffic, and in no
instance do we monitor or capture the behavior of other proxy users.
We do  not attempt to discover non-publicly listed
proxies by scanning the Internet.  By focusing exclusively on proxies
that are already publicly listed, we do not risk disclosing the
existence of any previously unknown proxy.

We achieve {\em Beneficience} by minimizing potential harms while
providing societal benefits (i.e., exposing the dangers of using open
proxies).  Unlike other studies that explicitly probe for instances of
Internet censorship by requesting potentially objectionable
content~\cite{weinberg2017topics,encore,encore-pubreview,echo-censorship},
our measurements avoid exposing proxy operators to risk (e.g.,
government sanction) by retrieving only URLs that are very unlikely to
be censored.  Nor do our measurements consume significant resources:
we request a small handful of URLs from each proxy; the average size
of the content is just 177~KiB.

Our study meets the Menlo Report's {\em Justice} criterion by
distributing our measurements equally across all identified open
proxies.

Finally, we achieve the {\em Respect for Law and Public Interest}
criterion by (i)~conducting only legal queries (i.e., we are not
requesting any content that is likely to be illegal or censored) and (ii)~being transparent
in our methods (see \S\ref{sec:overview}).

\Section{Conclusion}



Open proxies provide a free and simple way to bypass regional content
filters and achieve a limited degree of anonymity.  However, the
absence of any security guarantees for traffic passing through these
proxies makes their use highly risky: users may unintentionally expose
their traffic to malicious manipulations, especially when no
end-to-end security mechanisms (e.g., TLS) are present.

Our study of the Internet's open proxies---the largest conducted to date---discloses and quantifies new
forms of misbehavior, reinforcing the notion that open proxies should be
used with extreme caution.  We found numerous instances of misbehavior,
including the insertion of spurious ads and cryptocurrency-mining
Javascript, TLS MitM, and the injection of RATs
and other forms of malware.  Moreover, we found that
\confirmed{92\%} of advertised proxies listed on open proxy aggregator
sites are nonfunctional.  In contrast, our \confirmed{nearly monthlong} study
of the Tor network found zero instances of misbehavior and far greater
stability and goodput, indicating that Tor offers a safer and
more reliable form of proxied communication.

While we remain wary about the use of open proxies, some of the risks
we identify can be at least partially mitigated.  Tools such
as HTTPS Everywhere~\cite{https-everywhere} can help reduce the risk
of traffic manipulation by forcing end-to-end protections.  The
continued rollout of certificate transparency and similar measures
will also likely reduce (but not eliminate) risk, as they thwart the
certificate manipulation attacks described above.

We emphasize, however, that where e2e integrity and authenticity
guarantees are not possible (e.g., for unencrypted web traffic), the
use of open proxies still carries substantial risk.  Given users'
difficulty in adhering to safe browsing habits even in the absence of
proxies~\cite{felt2016-rethinking,warningland,cryingwolf}, we are
hesitant to recommend relying on browser-based protections to defend
against malicious proxy behavior.  Our findings suggest that the risks
of using open proxies are plentiful, and likely far outweigh their benefits.

\bibliographystyle{abbrvnat}
\setlength{\bibsep}{0pt}
\bibliography{micah-long.bib,other.bib}

\appendix

\section{Examples of HTTP Proxy Protocols}
\label{app:examples}

Example protocol interactions for HTTP and
CONNECT proxies are shown in Figures~\ref{fig:http}
and~\ref{fig:connect}, respectively.

\begin{figure}[h]
\small
  \centering
  {
    \color{blue}
\begin{Verbatim}[frame=single,commandchars=\\\{\}]
GET http://neverssl.com/index.html HTTP/1.1
User-Agent: Chrome/62.0.3202.94

\textcolor{red}{HTTP/1.1 200 OK}
\textcolor{red}{Date: Wed, 30 May 2018 14:28:02 GMT}
\textcolor{red}{Server: Apache}
\textcolor{red}{Content-Length: 12345}
\end{Verbatim}
}
\caption{Example HTTP requests (\textcolor{blue}{in blue}) and HTTP
  responses (\textcolor{red}{in red}) for HTTP proxies.}
\label{fig:http}
\end{figure}

\begin{figure}[h]
\small
{
\color{blue}
\begin{Verbatim}[frame=single,commandchars=\\\{\}]
CONNECT www.acsac.org:443 HTTP/1.1 
Host: www.acsac.org:443
Connection: keep-alive
User-Agent: Chrome/62.0.3202.94

\textcolor{red}{HTTP/1.1 200 Connection Established}
\textcolor{red}{Connection: close}
\end{Verbatim}
}
\caption{Example HTTP requests (\textcolor{blue}{in blue}) and HTTP
  responses (\textcolor{red}{in red}) for CONNECT proxies.}
\label{fig:connect}
\end{figure}

\section{Client locations}
\label{app:clients}

\begin{table}[h]
\centering
\begin{small}
\begin{tabular}{lc}
    \toprule
    {\bf Location} & {\bf Downtime} \\
    \midrule
California & --- \\
Canada & --- \\
Frankfurt & 1 day \\
Ireland & --- \\
London & --- \\
Mumbai & --- \\
Paris & --- \\
Ohio & --- \\
Oregon & --- \\
S\~{a}o Paulo & --- \\
Seoul & --- \\
Singapore & --- \\
Sydney & --- \\
Tokyo & --- \\
Virginia & --- \\
Author(s)' Institution {\em (blinded for review)} & 3 days \\
\midrule
Count: 16 & \\
\bottomrule
\end{tabular}
\caption{Locations of clients and the number of days during the
  measurement period in which that client did {\em not} participate.}
\label{tbl:clients}
\end{small}
\end{table}

\newpage
\section{File Manipulation Infections}
\label{app:file-manipulation}

\begin{table}[h]
\centering
\begin{small}
\begin{tabular}{lr}
\toprule
  {\bf Infection} & {\bf Frequency} \\
\midrule

W32/Behav-Heuristic-CorruptFile-EP          &       72.2\% (4781/6614) \\
W32.FamVT.ExpiroPC.PE                       &       63.7\% (4218/6614) \\
Heur.Corrupt.PE                             &       56.0\% (3704/6614) \\
malicious\_confidence\_100\% (D)              &       31.9\% (2111/6614) \\
static engine - malicious                   &       26.8\% (1777/6614) \\
Suspicious\_GEN.F47V0511                     &      23.2\% (1535/6614) \\
BehavesLike.Win32.Generic                   &       18.1\% (1200/6614) \\
malicious\_confidence\_90\% (D)               &      9.5\%  (632/6614) \\
Trojan.Crypt9                               &       7.6\% (503/6614) \\
Artemis                                     &       3.08\% (204/6614) \\

\bottomrule
\end{tabular}
\end{small}
\caption{Top 10 infections reported by VirusTotal for Windows executable response file type with unexpected content.}
\label{tbl:win32-top-infections}
\end{table}

\begin{table}[h]
\centering
\begin{small}
\begin{tabular}{lr}
\toprule
  {\bf Infection} & {\bf No. of Infected Files} \\
\midrule

HTML/ScrInject.B & 2471 \\
JS:Miner-Q & 2415 \\
Trojan.Script & 2283 \\
Trojan.ScrInject!8.A & 1158 \\

\bottomrule
\end{tabular}
\end{small}
\caption{Infections reported by VirusTotal for HTML response file type with unexpected content.}

\label{tbl:unexpected-html-infections}
\end{table}

\begin{table}[h]
\centering
\begin{small}
\begin{tabular}{lr}
\toprule
  {\bf Infection} & {\bf No. of Infected Files} \\
\midrule

Application.Bundler.iStartSurf & 545 \\
HEUR/AGEN.1004062 & 545 \\
Riskware.Win32.StartSurf.evlkna & 545 \\
SoftwareBundler:Win32/Prepscram & 545 \\
Trojan.Vittalia.13684 & 545 \\
W32/StartSurf.AU.gen!Eldorado & 545 \\
Win32:Evo-gen & 545 \\
heuristic & 545 \\
malware & 545 \\
not-a-virus:HEUR:AdWare.Win32.Generic & 545 \\
Application.Bundler.iStartSurf.DO & 545 \\
Application.Bundler.iStartSurf.DO (B) & 545 \\
Packed-UT!5694618272F0 & 520 \\
Packed-UT!4809C4235C2C & 25 \\

\bottomrule
\end{tabular}
\end{small}
\caption{Infections reported by VirusTotal for ISO response file type with unexpected content.}
\label{tbl:unexpected-iso-infections}
\end{table}

\begin{table}[t]
\centering
\begin{small}
\begin{tabular}{llr}
\toprule
    {\bf Unique Files} & {\bf Modification} & {\bf Count} \\
\midrule
1 & OC + HTTP Headers + Truncated OC & 1\\ 
2 & OC + 0 and Newline Character & 713 \\ 
3 & OC + </body></html> & 8 \\
4 & OC + 9 lines of harmless HTML & 298 \\
\midrule
Total &  & 1020 \\

\bottomrule
\end{tabular}
\end{small}
\caption{Unique responses for modified shell script. OC = Original Content}
\label{tbl:shell-script-results}
\end{table}

\end{document}